\documentclass[draft]{agujournal2019}

\usepackage{url} 
\usepackage{lineno}
\usepackage[inline]{trackchanges} 
\usepackage{soul}
\usepackage{amsmath,mathtools}
\usepackage{txfonts} 
\usepackage[utf8]{inputenc}
\usepackage[T1]{fontenc}

\usepackage{wasysym} 

\usepackage{rotating}
\usepackage{amssymb}    
\usepackage{cases}  
\usepackage{multirow} 
\usepackage{xcolor} 

\DeclareRobustCommand*{\deg}{\mathrm{^\circ}}

\usepackage{xfrac}
\usepackage{etoolbox}
 
\newcommand{\qty}[2]{$#1\,\textnormal{#2}$}

\usepackage{verbatimbox}
\usepackage{cprotect}

\renewcommand{\v}[1]{\ensuremath{\mathbf{#1}}} 
 
\newcommand{\pd}[2]{\frac{\partial #1}{\partial #2}} 
 
\let\baraccent=\= 
\renewcommand{\=}[1]{\stackrel{#1}{=}} 

\DeclareMathOperator{\arctantwo}{arctan2}

\newcommand*{\ditto}{\textquotedbl} 

\journalname{JGR: Space Physics}

\begin{document}

%
%

\title{A fast bow shock location predictor-estimator from 2D and 3D analytical models: Application to Mars and the MAVEN mission}

%
%




\authors{Cyril Simon Wedlund\affil{1}, Martin Volwerk\affil{1}, Arnaud Beth\affil{2}, Christian Mazelle\affil{3}, Christian M{\"o}stl\affil{1}, Jasper Halekas\affil{4}, Jacob Gruesbeck\affil{5}, and Diana Rojas-Castillo\affil{6}}


\affiliation{1}{Space Research Institute, Austrian Academy of Sciences, Graz, Austria}
\affiliation{2}{Department of Physics, Ume\aa\ University, 901\,87\ Ume\aa , Sweden}
\affiliation{3}{Institut de Recherche en Astrophysique et Planétologie (IRAP), Université de Toulouse, CNRS, UPS, CNES, Toulouse, France}
\affiliation{4}{Department of Physics and Astronomy, University of Iowa, Iowa City, IA, USA}
\affiliation{5}{NASA Goddard Space Flight Center, Laboratory for Planetary Magnetospheres, Greenbelt, MD, USA}
\affiliation{6}{Instituto de Geofísica, Universidad Nacional Autónoma de México, Coyoacán, Mexico} 





\correspondingauthor{C. Simon Wedlund}{cyril.simon.wedlund@gmail.com}




\begin{keypoints}
\item A simple predictor-corrector algorithm based on magnetic field data is presented to locate the bow shock position in spacecraft data. 
\item The method, biased towards quasi-perpendicular crossings, is general and applicable to all planetary bodies including Mars, Venus and Earth.
\item More than $14,900$ bow shock crossings are identified with MAVEN for Mars Years 32-35, with 2D/3D fits revealing North-South asymmetries. 
\end{keypoints}

%
%

%
%


\begin{abstract}
	We present fast algorithms to automatically estimate the statistical position of the bow shock from spacecraft data, using existing analytical two-dimensional (2D) and three-dimensional (3D) models of the shock surface. We derive expressions of the standoff distances in 2D and 3D and of the normal to the bow shock at any given point on it. 
	Two simple bow shock detection algorithms are constructed, one solely based on a geometrical predictor from existing models, the other using this predicted position to further refine it with the help of magnetometer data, an instrument flown on many planetary missions. 
	Both empirical techniques are applicable to any planetary environment with a defined shock structure. Applied to the Martian environment and the NASA/MAVEN mission, the predicted shock position is on average within $0.15$ planetary radius $R_p$ of the bow shock crossing. Using the predictor-corrector algorithm, this estimate is further refined to within a few minutes of the true crossing (\qty{\approx0.05}{$R_p$}). 
	Between 2014 and 2021, we detect $14,929$ clear bow shock crossings, predominantly quasi-perpendicular. Thanks to 2D conic and 3D quadratic fits, we investigate the variability of the shock surface with respect to Mars Years (MY), solar longitude (Ls) and solar EUV flux levels. Although asymmetry in $Y$ and $Z$ Mars Solar Orbital coordinates is on average small, we show that for MY32 and MY35, Ls = [135--225$\deg$] and high solar flux, it can become particularly noticeable, and is superimposed to the usual North-South asymmetry due in part to the presence of crustal magnetic fields.
\end{abstract}

\section*{Plain Language Summary}
[ enter your Plain Language Summary here or delete this section]

%
%

%


%
%
%
%

\section{Introduction}
Historically, planetary bow shocks, their position, size and shape, have been characterised statistically with the use of (empirical) analytical fitting models in two-dimensional (2D) or three-dimensional (3D) spatial coordinates. 
A classical starting point for characterising the Earth's bow shock in 3D includes the seminal work of \citeA{formisano_orientation_1979}, who investigated the asymmetry of the shock with respect to the apparent solar wind flow direction, with the use of quadratic surface fits with $9$ free parameters. In parallel, other studies such as that of \citeA{slavin_solar_1981} relied on a simple polar equation assuming axisymmetry along the Sun-planet line, corrected by the apparent motion of the solar wind in the rest frame of the planet, the so-called aberrated $X$ axis. The 2D approach has the merit of needing only $3$ free parameters but ignores the potential asymmetries of the shock as for example seen at Earth's bow shock \cite<e.g.>[]{formisano_orientation_1979,peredo_are_1993,peredo_three-dimensional_1995,merka_three-dimensional_2005}. 

More advanced physics-based models have also been proposed as a complement to those empirical attempts. A good introduction into analytical models of the bow shock, based on gasdynamic theory and magnetohydrodynamics solutions, is given in \citeA{verigin_planetary_2003} and recently in \citeA{kotova_physics-based_2021}. These studies present analytical functions describing the curvature, bluntness and skewing angle of the shock structure, which are arguably better suited to the fitting of the shock flanks; they are applicable to many planetary bow shock conditions.  

At Mars, due to the sparsity of early data and the non-collisional nature of the shock, the tendency has been to use in priority the simplest fitting model available with least free parameters, that is, an empirical 2D polar equation \cite{russell_relative_1977,slavin_solar_1991,trotignon_martian_2006,edberg_statistical_2008,hall_martian_2019}. Only recently with the NASA/Mars Atmospheric and Volatile EvolutioN (MAVEN) mission were quadratic fits used to characterise the general structure of the Martian bow shock, with \citeA{gruesbeck_three-dimensional_2018} providing fits to a careful subset of identified crossings in the first year of operations of the MAVEN mission. 

In recent years, many studies have attempted to characterise the Martian shock position and shape and its evolution under various solar wind and EUV conditions \cite<>[and references therein]{hall_martian_2019}. Two missions have been used for this goal, the ESA/Mars Express mission and the NASA/MAVEN mission. Mars Express (hereafter MEX for brevity) was launched in 2003 and has been orbiting Mars since 2004, whereas MAVEN was launched ten years later in November 2013 and has been orbiting the planet since 22 September 2014. MAVEN's scientific payload includes among others a fluxgate magnetometer \cite<MAG,>[]{connerney_maven_2015}, two ion spectrometers including the Solar Wind Ion Analyzer \cite<SWIA,>[]{halekas_solar_2015} and the Suprathermal and Thermal Ion Composition instrument \cite<STATIC,>[]{mcfadden_maven_2015}, and an electron spectrometer \cite<Solar Wind Electron Analyzer, SWEA,>[]{mitchell_maven_2016}. MEX unfortunately does not carry any magnetometer but includes a plasma suite (ions and electrons) as part of the ASPERA-3 package \cite{barabash_analyzer_2006}, which was used to investigate the plasma boundaries at Mars \cite{dubinin_plasma_2006}. Both missions aim at studying the upper atmosphere and the magnetospheric environment of Mars.

Detections of the bow shock in spacecraft data have relied on manual determinations using as many instruments (including plasma instruments and magnetometer) as available to avoid ambiguous detections \cite{gruesbeck_three-dimensional_2018}. Recently, \citeA{Nemec2020} proposed a region identification scheme based on selected plasma parameters and applied their technique to the MAVEN dataset in order to identify upstream solar wind, magnetosheath and magnetosphere regions crossed by the spacecraft. This method has the advantage of mapping these regions statistically, removing certain biases usually associated with manually picked individual boundary crossings which may be orbit-dependent. However, it requires the reliable knowledge of flow speed, ion density and magnetic field, which may not all always be available.
A parallel trend has also been to apply machine-learning techniques to plasma data for the labelling and identification of the regions crossed by a given spacecraft \cite<see>[for Mars and Earth, respectively]{hall_annual_2016, Breuillard2020} as part of online databases \cite{Genot2021}. However, these studies require extensive amounts of time and patience, as well as large computer resources to become efficient. Sometimes, a precise determination of the shock position in the data is not of paramount importance and simpler, faster, more straightforward approaches, such as the one presented in this study, can be advantageous. This may be the case in statistical studies where one of those regions needs to be systematically excluded, as in space weather databases monitoring the solar wind. This can also be of interest when areas around the predicted bow shock must be excluded for qualified reasons, as in wave studies focusing on regions outside of foreshock and shock wake structures, or when a first guess of the location and geometry of the shock is needed.

We present in this study new, simple analytical algorithms using two types of historical fitting techniques (2D and 3D), which make it possible to quickly estimate from spacecraft spatial coordinates the statistical geometrical position of the shock in planetary atmospheres. Special emphasis on the Martian environment and the MAVEN dataset is given throughout, however the method is applicable to any planetary environment and spacecraft dataset. This first crude estimator can be refined further by applying additional criteria, for example on the magnetic field amplitude measured by the MAVEN spacecraft. This provides a fast means to approximately and quite reliably identify the position of the shock so that solar wind and magnetosheath regions can be studied on a statistical level in the data. Moreover, other characteristics of the shock crossing, such as the quasi-parallel ($q_\parallel$) or quasi-perpendicular ($q_\perp$) geometry of the shock can easily be obtained by deriving the perpendicular direction to the shock at any point on the surface. 

After a review of 2D and 3D bow shock fitting models at Mars in Section\ \ref{sec:bowshockModels}, we present in a consistent manner the leading equations behind these models in 2D and 3D, give the analytical expressions for standoff distances, and propose a geometric calculation of the normal to the bow shock at any point on the surface. Starting in Section\ \ref{sec:bowshockDetection}, we introduce a predictor algorithm for a fast estimation of the shock position in spacecraft orbital coordinates and its timing (Section\ \ref{sec:Predictor}). In Section\ \ref{sec:PredictorCorrector}, we propose a simple correction on this position and timing with the sole help of magnetometer data (predictor-corrector algorithm). Application to the MAVEN MAG dataset is then given as validation on a few examples and then extended to the whole available dataset. Finally, as a result of the automatic detection proposed here, statistical analytical fits are given for the MAVEN mission between November 2014 and February 2021, with a discussion of the shock's asymmetry based on terminator and standoff distances. Applications for space weather-related databases are also mentioned.

\section{Bow shock models at Mars}\label{sec:bowshockModels}

In this section, following a survey of existing fitting models at Mars, we present comprehensive formulae for analytical fits in 2D polar coordinates and 3D Cartesian coordinates, with a calculation of bow shock subsolar and terminator standoff distances. We also show how to calculate the normal to the surface at a given point in space, in order to estimate the $q_\perp-q_\parallel$ shock conditions.

\subsection{Coordinate systems and solar wind flow aberration}\label{sec:aberration}
All spacecraft coordinates in this study are in Mars Solar Orbital coordinates (MSO) for simplicity, in accordance with most previous studies. In the MSO system, identical to the Sun-state coordinate system, the $+X_\text{MSO}$ axis points towards the Sun from the planet's centre, $+Z_\text{MSO}$ is towards Mars' North pole and perpendicular to the orbital plane defined as the $X_\text{MSO}$--$Y_\text{MSO}$ plane passing through the centre of Mars, and $Y_\text{MSO}$ completes the orthogonal system.

\begin{figure}[t]
 \includegraphics[width=\columnwidth]{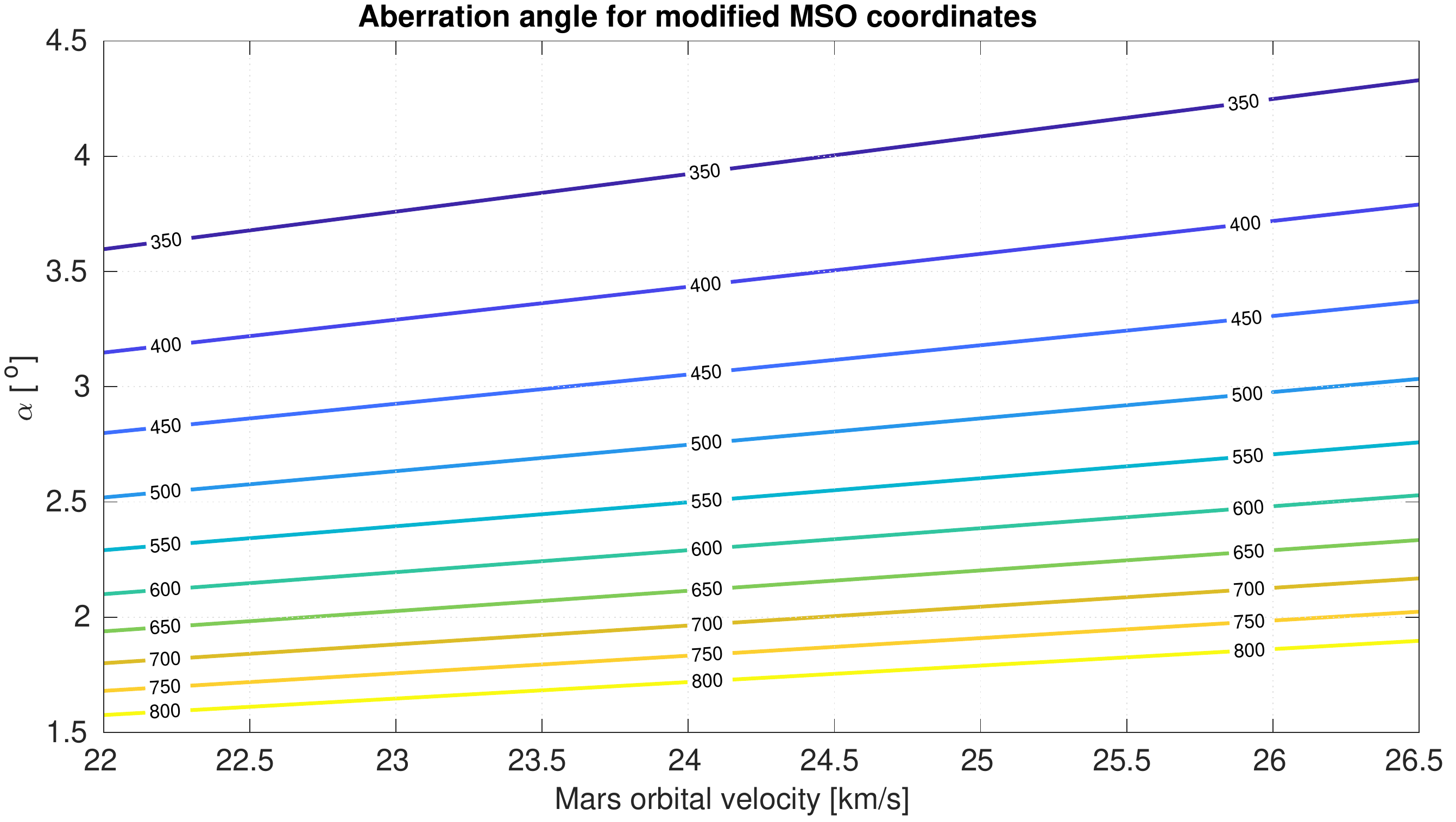}
 \caption{Aberration angle with respect to the orbital velocity of Mars (in km/s) and the solar wind mean speed (colour-coded isocontours, in km/s). \label{fig:aberrationAngle} }
\end{figure}

Because of the orbital motion of Mars with respect to the average direction of the solar wind, the apparent direction of the  solar wind in the rest frame of the planet deviates from the anti-sunward direction. As a result, an {\it anti-clockwise} rotation by an angle $\alpha$ around the $Z$ axis must be applied so that the bow shock's major axis is aligned with respect to the $X$ axis along the solar wind flow. This aberration, first seen in cometary tails and at the origin of the hypothesis by Biermann of a stellar wind \cite{biermann_kometenschweife_1951}, is taken into account in the so-called {\it aberrated MSO coordinates}, denoted $X^\prime_\textrm{MSO}$, $Y^\prime_\textrm{MSO}$ and $Z^\prime_\textrm{MSO}$ (although $Z$ is left unchanged by the transformation). To unclutter notations, the 'MSO' subscript is now dropped. Following \citeA{formisano_three-dimensional_1979}, \citeA{slavin_solar_1981} define the angle $\alpha$ as $\alpha = \tan^{-1}\left(V_\textrm{p}/\varv_\textrm{sw}\right)$
where $V_\textrm{p}$ is Mars' orbital velocity and $\varv_\textrm{sw}$ is the solar wind velocity, for example expressed in km\,s$^{-1}$.
Mars' average orbital velocity is $V_\textrm{p} = 24.07_{22.0}^{26.5}$~km\,s$^{-1}$. For the maximum value ($26.5$~km\,s$^{-1}$), the angle 
is $\alpha = 3.8\deg$ for a typical solar wind speed of \qty{400}{km\,s$^{-1}$}. The angle assumed by all studies except those of \citeA{slavin_solar_1981} and \citeA{slavin_solar_1991} is $4\deg$. In \citeA{slavin_solar_1981}, the aberration angle was chosen to be varying with solar wind speed conditions. In \citeA{slavin_solar_1991}, $\alpha = 3.2\deg$. Figure\ \ref{fig:aberrationAngle} shows the aberration angle with respect to orbital velocity (abscissa) and to solar wind velocity (colour code).

\subsection{Parametric models}\label{sec:paramModels}

Table\ \ref{tab:BowShockSolarActivity} chronologically lists past bow shock studies at Mars and their characteristics in terms of solar activity, solar cycle number and number of observations, including the recent MEX and MAVEN surveys. 

Bow shock models at Mars have been proposed since the end of the 1970s, including (but not limited to): \citeA{russell_relative_1977, slavin_solar_1981, slavin_solar_1991, schwingenschuh_martian_1990, trotignon_location_1991, zhang_asymmetries_1991, zhang_interplanetary_1991, trotignon_position_1993, Verigin1999, vignes_solar_2000, trotignon_martian_2006, edberg_statistical_2008, edberg_magnetosonic_2010}. These studies were performed with several spacecraft including Viking, Mars Global Surveyor (MGS) and Phobos-2, and for varying solar conditions. 
\citeA{gringauz_electron_1976} quoted in \citeA{russell_relative_1977} reported $11$ crossings for the Russian Mars-2,3,5 satellites, but \citeA{slavin_solar_1981} later reanalysed the datasets and found $14$ crossings in total. \citeA{slavin_solar_1991} reported $94$ crossings for Phobos-2, upped to $126$ by \citeA{trotignon_position_1993}. Using the same dataset, \citeA{Verigin1993} reported a weak dependence of terminator bow shock position on solar wind dynamic pressure $P_\text{sw}$; additionally in an analogy with magnetised planets regarding the dependence of the magnetopause thickness to $P_\text{sw}$, these authors anticipated the discovery of crustal magnetic field sources at Mars, later vindicated with MGS measurements \cite{Acuna1998}. 

In contrast with what was found at Venus, \citeA{slavin_solar_1981} and later \citeA{vignes_solar_2000,vignes_factors_2002} (using MGS and data from previous missions) suggested that the mean bow shock standoff distance was likely independent of the solar activity. \citeA{slavin_solar_1991} showed that the terminator distance, which is a marker of the swelling of the cavity flanks, varied by as much as $11\%$ between the Mars-2,3,5 observations (low activity) and the Phobos-2 observations (high activity), although the number of crossings for each mission largely differed. 
Mars EXpress (MEX), with its very long activity spanning the end of solar cycle $23$ and cycle $24$ up to now (towards the beginning of new cycle $25$), had the best chance to conclusively solve this aspect: \citeA{hall_martian_2019} found that for the years $2004$--$2017$, the terminator distance underwent variations up to $\sim7\%$, in agreement with the preliminary results of \citeA{trotignon_position_1993}. Of note, \citeA{mazelle_bow_2004} made a review of all available measurements before MEX started observing, and discussed the solar cycle variations and the differences observed with Venus \cite<for which up to $35\%$ increases of the bow shock location in the terminator plane with increasing activity have been reported, see>[]{russell_solar_1988,zhang_initial_2008,edberg_magnetosonic_2010}.

Of great import, \citeA{edberg_plasma_2009,edberg_magnetosonic_2010} used MGS and MEX data in combination with ACE data extrapolated to Mars to study the dependence of the bow shock location to solar EUV flux and magnetosonic Mach number (noted $M_\text{ms}$). They pointed out that the shape of the magnetosonic shock wave depends on the ratio of the solar wind speed to the magnetosonic speed. Later, for the entire period $2004$--$2015$, \citeA{hall_annual_2016} identified $11,861$ crossings in the MEX database, using electron spectrometer data with machine-learning algorithms. This study was extended by \citeA{hall_martian_2019} up to $2017$ with a total of $13,585$ crossings, totalling 13 years of operations of MEX. Both works used a standard 2D conic fit depending on the Mars Year (MY), with observed variations up to a few percent in terms of standoff bow shock distance. In addition to solar wind cycle variations and hemispherical changes from a 2D perspective, these studies confirmed the dependence of the shock position with $P_\text{sw}$ and, most drastically, with solar EUV flux. Correspondingly, \citeA{ramstad_solar_2017} studied with coincident electron and ion data a subset of only $1,083$ inbound and outbound MEX orbits for the period $2005-2016$. They evaluated the dependence of the Induced Magnetospheric Boundary (IMB) and bow shock (BS) to EUV flux and the solar wind's lowest moments (density, bulk velocity), showing that the BS mostly expands and contracts with the IMB. However, they also found that the BS swelling in the flank due to increased EUV fluxes cannot be solely explained by a corresponding swelling of the IMB. Simultaneously with MAVEN (both magnetometer and ion measurements), \citeA{halekas_structure_2017} investigated how the Martian magnetosphere and bow shock responded to EUV flux, $M_\text{ms}$ and $P_\text{sw}$ between October 2014 and May 2016 ($0.85$ Mars year). In agreement with previous studies, they showed that the shock inflates with increasing EUV flux and contracts with increasing dynamic pressure and $M_\text{ms}$; this in turn leads to EUV flux and dynamic pressure competing against one another because of their common $1/d_h^2$ dependence on heliocentric distance $d_h$.

Recently, \citeA{Nemec2020} used MAVEN plasma and magnetometer data to construct maps of the solar wind/magnetosheath regions. Their modelled bow shock locations, explicitly dependent on $P_\text{sw}$, EUV flux and crustal field intensities, were in good agreement with the average fits of \citeA{trotignon_martian_2006}, with appreciable differences in terminator extensions as compared to the results of \citeA{ramstad_solar_2017}.

\begin{table}[t]
    \centering
    \caption{Statistical studies on the Martian bow shock position replaced chronologically (with respect to in-situ observations) in the context of solar activity and Mars Year (MY). N is the number of bow shock crossings considered in each study. MGS = Mars Global Surveyor. MEX = Mars Express.}
    \label{tab:BowShockSolarActivity}
    \scriptsize
    \begin{tabular}{l | c c r | c c c c | c }
        {\bf Reference}           & {\bf Spacecraft}            & {\bf Years} & {\bf N} & {\bf Solar activity} & {\bf Cycle \#}  & \bf{Start} & {\bf Max.} &  {\bf MY} \\
        \hline
        \citeA{slavin_solar_1981} & Mariner\,4                   & $1965$      & $2$   & \emph{Low} & $20$ & $1964$ & $1968$ & $6$ \\
        \citeA{russell_relative_1977}$^{a}$ & Mars\,2, 3, 5                      & $1965-1974$      & $11$   & \emph{Low-Medium}    & \ditto & \ditto & \ditto  & $9-11$ \\
        \citeA{slavin_solar_1981} & Mars\,2, 3                   & $1971-1972$ & $10$  & \emph{Medium} & \ditto & \ditto & \ditto  & $9-10$ \\
        \citeA{slavin_solar_1981} & Mars\,5                      & $1974$      & $4$   & \emph{Low}    & \ditto & \ditto & \ditto  & $11$ \\
        \citeA{slavin_solar_1991} & Mariner\,4, Mars\,2, 3, 5                      & $1965-1974$      & $24$   & \emph{Low-Medium}    & \ditto & \ditto & \ditto  & $6-11$ \\
        \citeA{schwingenschuh_martian_1990} & Phobos\,2                    & $1989$      & $\sim100$  & \emph{High}  & $22$ & $1986$ & $1989$ & $19$ \\
        \citeA{slavin_solar_1991} & \ditto                    & \ditto      & $94$  & \emph{High}  & $22$ & \ditto & \ditto & \ditto \\
        \citeA{trotignon_position_1993} & \ditto & \ditto & $26$ & \ditto & \ditto & \ditto & \ditto  & \ditto\\
        \citeA{trotignon_position_1993}$^{b}$ & \ditto & \ditto & $126$ & \ditto & \ditto & \ditto & \ditto  & \ditto\\
        \citeA{Verigin1993}$^{c}$                     & \ditto & \ditto & \ditto  & \ditto & \ditto & \ditto & \ditto  & \ditto\\
        \citeA{Verigin1999}                               & \ditto & \ditto & \ditto  & \ditto & \ditto & \ditto & \ditto  & \ditto\\
        \citeA{vignes_solar_2000} & MGS                         & $1997-1998$ & $450$ & \emph{Low}  & $23$ & $1996$ & $2001$ & $23-24$ \\
        \citeA{trotignon_martian_2006} & MGS & $09/1997-02/1999$ & $573$ & \emph{Low-Medium} & \ditto & \ditto & \ditto  & \ditto\\
        \citeA{edberg_statistical_2008} & MGS                   & \ditto & $619$ & \ditto & \ditto & \ditto & \ditto  & \ditto \\
        \citeA{hall_annual_2016,hall_martian_2019}$^{d}$  & MEX                & $2004-2008$ & $4,422$ & \emph{Medium-Low} & \ditto & \ditto & \ditto  & $27-29$ \\
        \citeA{ramstad_solar_2017}$^{e}$       & MEX                & $11/2005-12/2016$ & $2,166$ &  \emph{High-Medium} & $23, 24$ & $1996, 2008$ & $2001, 2014$ & $27-33$ \\  
        \citeA{hall_annual_2016,hall_martian_2019}$^{f}$  & MEX                & $2008-2015$ & $7,669$ & \emph{Low-High}  & $24$ & $2008$ & $2014$  & $30-32$ \\
        \citeA{hall_martian_2019}$^{g}$ & MEX                & $2015-12/2017$ & $1,494$ & \emph{High-Medium}  & \ditto & \ditto & \ditto  & $33$ \\
        \citeA{halekas_structure_2017}$^{h}$ & MAVEN               & $10/2014-05/2016$ & $-$ & \emph{High-Medium}  & \ditto & \ditto & \ditto  & $32-33$ \\
        \citeA{gruesbeck_three-dimensional_2018} & MAVEN        & $11/2014-04/2017$ & $1,799$ & \emph{High-Medium}  & \ditto & \ditto & \ditto  & $32-34$ \\
        \citeA{Nemec2020}$^{i}$ & MAVEN        & $11/2014-02/2020$ & $-$ & \emph{High-low}  & \ditto & \ditto & \ditto  & $32-35$ \\        
    \hline
    \multicolumn{9}{l}{$^{a}$Observations by \citeA{gringauz_electron_1976} and analysed further by \citeA{russell_relative_1977}.}\\ 
    \multicolumn{9}{l}{$^{b}$In \citeA{trotignon_martian_2006}, $127$ Phobos\,2 crossings of the bow shock were reported, that is, one more than in \citeA{trotignon_position_1993}.}\\
    \multicolumn{9}{l}{$^{c}$Number of observations from Phobos-2 used in this study was presumably the same as in \citeA{trotignon_position_1993}.}\\  
    \multicolumn{9}{l}{$^{d}$Data from \citeA{hall_martian_2019}, MY27-29, Table~3 with respect to Mars Years. MY29 runs from 09-12-2007 to 25-10-2009, hence overlapping}\\
    \multicolumn{9}{l}{\hspace{2ex} slightly with Solar Cycle 24, although still at minimum level of activity.}\\  
    \multicolumn{9}{l}{$^{e}$About $7,000$ orbits were first manually examined, ``out of which $1,083$ orbit inbound and outbound segments with identified BS, IMB [...]}\\
    \multicolumn{9}{l}{\hspace{2ex} crossings were included.'' Orbital coverage of MEX is shown in their Figure\ 9. No discrimination with solar cycle or Mars Year}\\
    \multicolumn{9}{l}{\hspace{2ex} is given, although EUV flux and solar-wind parameter dependence are studied.}\\  
    \multicolumn{9}{l}{$^{f}$Data from \citeA{hall_martian_2019}, MY30-32, Table~3 with respect to Mars Years. MY32 runs from 31-07-2013 to 17-06-2015.}\\
    \multicolumn{9}{l}{$^{g}$Data from \citeA{hall_martian_2019}, MY33, Table~3 with respect to Mars Years. MY33 runs from 18-06-2015 to 04-05-2017.}\\
    \multicolumn{9}{l}{$^{h}$Bow shock variations are obtained by fitting 2D-gridded datasets of average plasma density jumps through the shock location as measured with}\\
    \multicolumn{9}{l}{\hspace{2ex} MAVEN/SWIA and are discriminated against magnetosonic Mach number $M_{\rm ms}$, EUV flux and dynamic pressure.}\\
    \multicolumn{9}{l}{$^{i}$Number of individual crossings not disclosed due to the nature of the region detection scheme used. $2,040$ full orbits reported.}
    \end{tabular}
\end{table}

Broadly speaking, two approaches fitting the shape of bow shocks have historically been employed, one using a simple 2D polar form \cite<e.g.>[]{slavin_solar_1981}, the other the 3D general Cartesian conic form \cite<e.g.>[]{formisano_three-dimensional_1979,formisano_orientation_1979}. Semi-empirical models based on gasdynamic and MHD predictions such as those of \citeA{Verigin1993,Verigin1999} are not discussed in the following.

\subsubsection{2D polar form}

\begin{figure}[t]
 \includegraphics[width=\columnwidth]{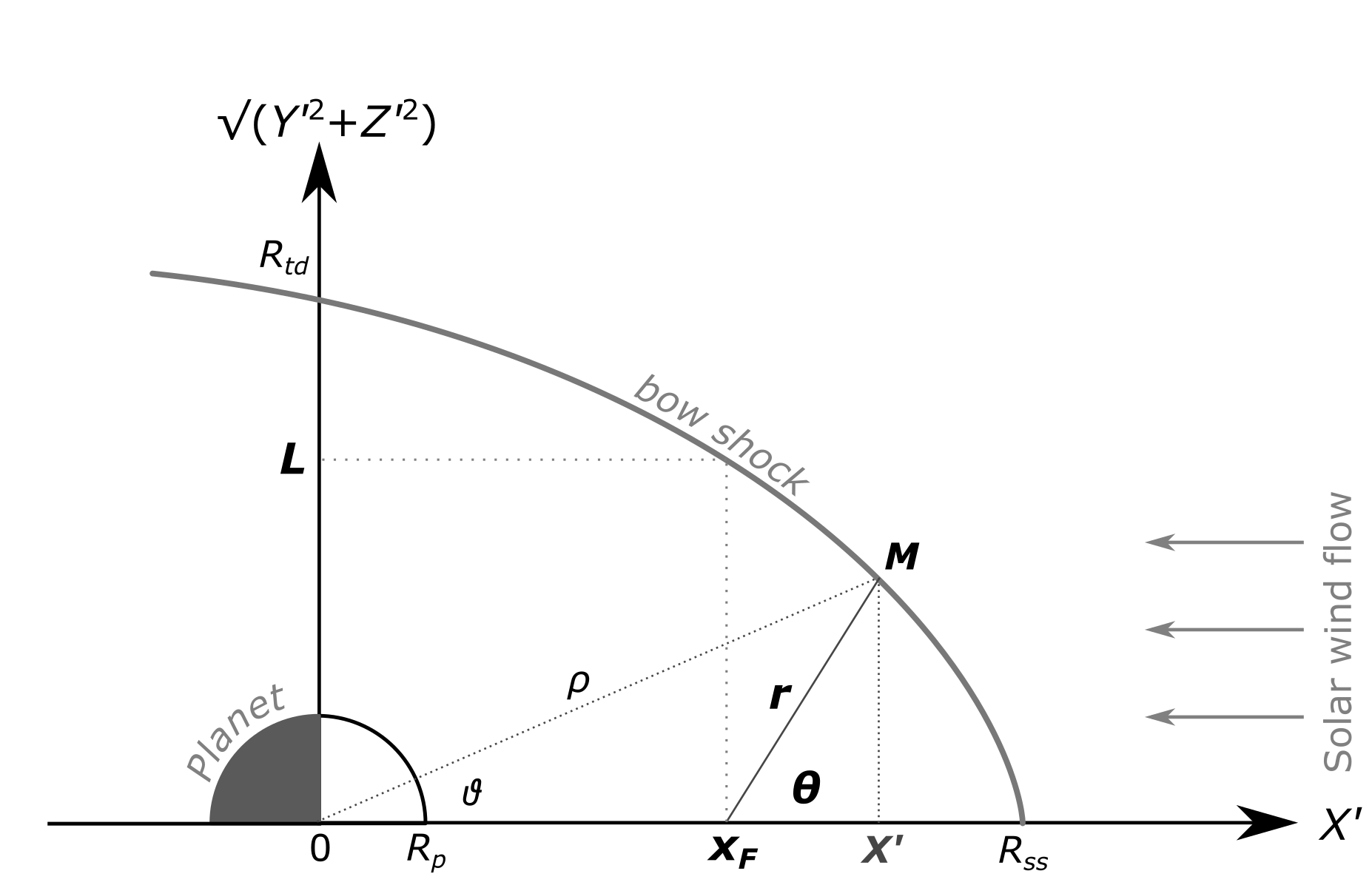}
 \caption{Typical 2D conic bow shock shape in the aberrated $(X^\prime,\sqrt{Y^{\prime 2}+Z^{\prime 2}})$ coordinate system.  For a point $M$ on the shock surface, $\rho$ is the Euclidean distance to the shock from the centre of the planet of radius $R_\textrm{p}$, and $r$ is the distance to the shock surface from the focus $x_F$ of the conic with semilatus rectum $L$ and making an angle $\theta$ with the $X^\prime$ direction, so that Equation\,(\ref{eq:polarBSmodel}) holds. $\vartheta$ is the usual polar angle, with respect to the centre of Mars. $R_\text{ss}$ and $R_\text{td}$ are the standoff subsolar and terminator distances.  \label{fig:2dconic} }
\end{figure}

Assuming that a full-formed bow shock in aberrated coordinates is symmetric with respect to the $X^{\prime}$ axis, the 3D shape of the bow shock can be reduced to a 2D problem in the $(X^{\prime}, \sqrt{Y^{\prime 2}+Z^{\prime 2}})$ plane. All Martian studies except that of \citeA{gruesbeck_three-dimensional_2018} have made this assumption. A simple 2D conic of revolution (usually a parabola or a hyperbola), symmetric around the aberrated MSO axis $X^{\prime}$ and decentred from its focus $x_F$ is shown in Figure\ \ref{fig:2dconic}. Such a 2D conic takes the parametric form \cite<for example,>[]{hall_martian_2019}:
\begin{linenomath}
\begin{align}
    &r = \frac{L}{1 + \epsilon \cos\theta},\label{eq:polarBSmodel}\\
    \textrm{with: }& r = \sqrt{(X^{\prime}-x_F)^2+Y^{\prime\,2}+Z^{\prime\,2}},\\
    \textrm{and: }& \cos \theta = \frac{X^{\prime} - x_F}{r}
\end{align}{}
\end{linenomath}
where $\epsilon$ is the conic's eccentricity,  $L$ the semilatus rectum \cite<called \emph{terminator crossing} by>[at Venus, because the focus is taken at the centre of the planet]{zhang_initial_2008,volwerk_mirror_2016} and $\theta$ the angle measured from the focus of the conic, typically within the $[-\pi/2, \pi/2]$ range. This range of angles depends on the nature of the conic section: for a parabola ($\epsilon = 1$) $\theta \in ]-\pi, \pi[$ (borders excluded), for an ellipse ($\epsilon < 1$) $\theta \in ]-\pi, \pi]$, and for a hyperbola ($\epsilon > 1$) $\exists\,\theta_0 \in ]0,\pi/2[\ \ \mid\ \cos \theta_0 = 1/\epsilon$, and $\theta \in ]-\theta_0,\theta_0[\ \cup\ ]\theta_0,2\pi-\theta_0[$. 
The equivalent rectangular (Cartesian) form of this equation is \cite{trotignon_martian_2006}:
\begin{linenomath}
\begin{align}
    Y^{\prime\,2} + Z^{\prime\,2} - \left(\epsilon^2-1\right) \left(X^\prime-x_F\right)^2 + 2\epsilon L \left(X^{\prime}-x_F\right) - L^2 = 0.
    \label{eq:cartBSmodel2D}
\end{align}{}
\end{linenomath}
In this representation, \citeA{trotignon_martian_2006} derived two additional useful quantities, the standoff shock distance along the $X$ axis, $R_\mathrm{ss}$ \cite<also called subsolar aerocentric distance in>[]{trotignon_martian_2006} and the standoff terminator distance $R_\mathrm{td}$ along the cylindrical coordinate $\sqrt{Y^{'2}+Z^{'2}}$ \cite<which is none other than the diameter of the tail at $X^\prime = 0$ divided by 2, or as in>[ the ``dawn radius'']{russell_relative_1977}:
\begin{linenomath}
\begin{align}
    R_\mathrm{ss} &= x_F + \frac{L}{1+\epsilon}, \label{eq:standoffDistanceConic} \\
    R_\mathrm{td} &= \sqrt{L^2 + (\epsilon^2 - 1)\,x_F^2 + 2\epsilon\, L\, x_F}. \label{eq:standoffTerminatorDistanceConic}
\end{align}{}
\end{linenomath}
We can derive another parameter of interest, that is, the aperture of the Mach cone related to the shock structure --the {\it limiting Mach cone angle} \cite{Verigin2003MachCone}. In the gasdynamic approach, it is defined as $\varrho = \arcsin\left(1/M_\textrm{s}\right)$, where $M_\textrm{s} ={\varv_\text{SW}}/\varv_\textrm{s}$ is the sonic Mach number. The sonic speed is $\varv_\textrm{s} = \sqrt{\gamma P/\rho}$, with $\rho$ the solar wind ion mass density, $\gamma = 5/3$ the ratio of specific heats, and $P = n_\textrm{sw}\,k_B (T_e + T_i)$ the solar wind thermal pressure, with $T_e$ and $T_i$ the electron and ion temperatures, respectively. For a hyperbola ($\epsilon$\,$>$\,$1$), the limiting Mach cone angle is exactly the angle made by the asymptotes of the hyperbola \cite{slavin_planetary_1984}:
\begin{linenomath}
\begin{align}
    \varrho &=\tan^{-1}\sqrt{\epsilon^2-1}\label{eq:MachConeAngle}\\
    \text{with}: \Delta\varrho &= \frac{\Delta\epsilon/\epsilon}{\sqrt{\epsilon^2-1}}
\end{align}
\end{linenomath}
as uncertainty. In a canonical form for the hyperbola, with $a$ the distance from the nose to the intersection of the asymptotes on the $X^\prime$ axis, and $b$ that from the shock nose to the asymptote on the $Y^\prime$ axis, $\arctan\left(b/a\right)$. Since, by definition $\epsilon = \sqrt{1+b^2/a^2}$, the substitution readily yields expression\,(\ref{eq:MachConeAngle}). It is noteworthy to remark that for $\epsilon$ close to $1$, the uncertainty increases to infinity; any determination of $\varrho$ is thus unreliable for quasi-parabolic curves.

For the fit, \citeA{slavin_solar_1981} and \citeA{slavin_solar_1991} rewrote Equation\,(\ref{eq:polarBSmodel}) as $y = ax + b$ (posing $y=1/r$, $x=\cos\theta$, $a=\epsilon/L$ and $b=1/L$) and performed simple linear regressions for a range of foci locations. As pointed out by \citeA{vignes_solar_2000}, this may result in fitting biases when observations are widely disparate in their location: in this case direct fitting methods to Equation\,(\ref{eq:polarBSmodel}) should be preferred.
With a direct polar fit to MGS data, \citeA{edberg_statistical_2008} gave for example the following fitted values: $\epsilon = 1.05\pm0.04$, $L   = 2.10\pm0.09$, $x_F = 0.55\pm0.12$. However, to match the results plotted in Figure\ 1 of \citeA{edberg_statistical_2008}, the value of $\epsilon$ must be modified down to $\epsilon = 1.03$, a marginal difference likely due to rounding errors. For comparison, the corresponding values derived by \citeA{hall_martian_2019}, for MEX data but with a larger sample, are $\epsilon = 0.998\pm0.003$, $L = 1.802\pm0.002$ and $x_F = 0.76$. For MGS and MEX data \cite{edberg_statistical_2008,hall_annual_2016,hall_martian_2019}, the subsolar standoff distance is $R_\textrm{ss} = 1.63\pm0.04\,R_\textrm{p}$, whereas the terminator standoff  distance is $R_\textrm{td} = 2.50\pm0.09\,R_\textrm{p}$.
By comparison, \citeA{halekas_structure_2017} found large variations of the bow shock standoff distances in the early MAVEN data, with $R_\textrm{ss}\sim 1.6$--$1.9\,R_\textrm{p}$ and $R_\textrm{td}\sim 2.5$--$3.1\,R_\textrm{p}$ depending on EUV flux levels and combined with either $M_\text{ms}$ or the solar wind dynamic pressure. Differences towards upper values with previous studies likely stem from different EUV levels encountered by the respective missions, the solar EUV flux being one of the main drivers, through ionosphere and exosphere variations, of the bow shock position \cite{halekas_structure_2017,hall_martian_2019}.

We present in Table\ \ref{tab:Conic2DparametersPast} the fitted conic parameters of the main quoted references in Table\ \ref{tab:BowShockSolarActivity}, in chronological order. It is interesting to remark that most shapes fitted are \emph{stricto sensu} hyperbolic ($\epsilon \geq 1$), but in practice can be considered quasi-parabolic as eccentricity $\epsilon \sim 1$, which makes it possible to calculate the limiting Mach cone angle. For example, when fitting bow shocks from different MYs, \citeA{hall_martian_2019} showed that eccentricities varied around $\epsilon = 1$ by less than $5\%$ between MY27 and MY33, with a marked tendency towards ellipsoidal shapes (only $2$ consecutive years, MY28 and MY29, had eccentricities above $1$).

\begin{table*}[t]
    \centering
    \caption{Summary of Martian bow shock 2D conic parameters. Pre-Mars Express results were already summarised in \citeA{trotignon_martian_2006}, Table\ 1. The aberration angle $\alpha$ is given for each reference. $\varrho$ is the limiting Mach cone angle, calculated by formula\,(\ref{eq:MachConeAngle}) in the case of a hyperbolic shape. The mean value for each mission is also given, with Mars 2-3-5 and Mariner 4 \cite{slavin_solar_1991}, Phobos 2 \cite{trotignon_position_1993}, Mars Global Surveyor (MGS) \cite{edberg_statistical_2008} and Mars Express (MEX) \cite<>[]{hall_annual_2016,hall_martian_2019}. The planetary radius of Mars is by definition $R_\textrm{p} = 3389.5$\,km.
    }
    \label{tab:Conic2DparametersPast}
    \scriptsize
    \begin{tabular}{l | c c c c c c | l c}
        {\bf Reference} & $\epsilon$ & $L$ [$R_\textrm{p}$] & $x_F$ [$R_\textrm{p}$] & $R_\text{ss}$ [$R_\textrm{p}$] & $R_\text{td}$ [$R_\textrm{p}$] & $\alpha$ & Nature & $\varrho$ [$\deg$]\\
        \hline
        \citeA{russell_relative_1977}$^{a}$    & $0.99\pm0.11$ & $2.985$ & $0$ & $1.50\pm0.15$ & $3.00\pm0.13$  & $0$ & Ellipse & $-$\\
        \citeA{slavin_solar_1981}$^{b}$         & $0.94\pm0.04$ & $1.94\pm0.02$ & $0.5$ & $1.50\pm0.04$ & $2.36$$^{m}$  & $\tan^{-1}\frac{V_p}{V_\textrm{sw}}$ & Ellipse & $-$\\
        \citeA{slavin_solar_1991}$^{c}$       & $1.02$ & $1.68$ & $0.7$ & $1.55$ & $2.29$$^{m}$  & $\tan^{-1}\frac{V_p}{V_\textrm{sw}}$ & Hyperbola & $11.4\pm2.9$\\
        \citeA{schwingenschuh_martian_1990} & $0.85$ & $2.72$ & $0$ & $1.47\pm0.03$ & $2.72$$^{m}$ & $3.8\deg$ &  Ellipse & $-$\\
        \citeA{trotignon_location_1991} & $0.95\pm0.10$ & $2.17\pm0.03$ & $0.5$ & $1.62\pm0.07$ & $2.60$$^{m}$ & $4\deg$ &  Ellipse & $-$\\
        \citeA{trotignon_position_1993} & $1.02\pm0.01$ & $2.17\pm0.03$ & $0.5$ & $1.57\pm0.03$ & $2.6$ & $4\deg$ &  Hyperbola & $11.4\pm2.8$\\
        \citeA{vignes_solar_2000}$^{d}$       & $1.03\pm0.01$ & $2.04\pm0.02$ & $0.64\pm0.02$ & $1.64\pm0.08$ & $2.62\pm0.09$ & $4\deg$ & Hyperbola & $13.9\pm2.3$\\
        \citeA{vignes_solar_2000}$^{e}$        & $1.02\pm0.02$ & $1.93\pm0.01$ & $0.72$ & $1.67\pm0.03$ & $2.56\pm0.06$ & $4\deg$ & Hyperbola & $11.4\pm5.6$\\
        \citeA{trotignon_martian_2006}  & $1.026\pm0.002$ & $2.081\pm0.006$ & $0.6$ & $1.63\pm0.01$ & $2.63\pm0.01$ & $4\deg$ & Hyperbola & $12.9\pm0.5$\\
        \citeA{edberg_statistical_2008}$^{f}$ & $1.05\pm0.04$& $2.10\pm0.09$ & $0.55\pm0.12$ & $1.58\pm0.18$ & $2.69$$^{m}$  & $4\deg$ & Hyperbola & $17.8\pm6.8$\\
        \citeA{hall_annual_2016} & $1.01\pm0.11$ & $1.82\pm0.08$ & $0.74\substack{+0.03 \\ -0.10}$ & $1.65\substack{+0.13 \\ -0.18}$ & $2.46\substack{+0.20 \\ -0.22}$ & $4\deg$ & Hyperbola & $-$$^{k}$\\
        \citeA{halekas_structure_2017}$^{g}$   & $1.0$ & $2.01$--$2.54$ & $0.6$ & $1.6$--$1.9$$^{l}$ & $2.5$--$3.1$$^{m}$ & $4\deg$ & Parabola & $-$\\ 
        \citeA{ramstad_solar_2017}$^{h}$   & $1.022$ & $1.48$ & $0.85$ & $1.58$ & $2.19$ & ($4\deg$) & Hyperbola & $11.9\pm2.7$\\ 
        \citeA{hall_martian_2019}$^{i}$ & $0.998\pm0.001$ & $1.802\pm0.002$ & $0.76$ & $1.662$$^{l}$ & $2.445\pm0.003$ & $4\deg$ & Ellipse & $-$\\
        \hline
        All (one per mission)$^{j}$ & $1.016\pm 0.012$ & $2.01\pm 0.25$ & $0.61\pm0.10$ & $1.61\pm 0.08$ & $2.56\pm 0.20$ & $4\deg$ & Hyperbola & $13\pm 4$ \\
        \hline       
        \multicolumn{9}{l}{$^{a}$Because the Mars\,2, 3 and 5 measurements reported by \citeA{gringauz_electron_1976} (in total $11$ crossings) did not specify local times, aberration angle $\alpha$ was assumed}\\
        \multicolumn{9}{l}{\hspace{2ex} to be zero.}\\
        \multicolumn{9}{l}{$^{b}$These authors use the full definition of the aberration angle, resulting in $\alpha = \arctan\left(V_p/\varv_\textrm{sw}\right)$, in contrast to the more recent studies. See Section~\ref{sec:aberration}.}\\
        \multicolumn{9}{l}{$^{c}$Mariner\,4, and Mars\,2, 3, 5 data only here. Uncertainties on $\epsilon$ fitted values assumed to be $1\%$.} \\
        \multicolumn{9}{l}{$^{d}$"Direct fit" method with all $3$ parameters varying simultaneously.} \\
        \multicolumn{9}{l}{$^{e}$"Slavin's method", using a linear regression in ($1/r$,$\cos\theta$) space.}\\
        \multicolumn{9}{l}{$^{f}$Note that $\epsilon = 1.03$ matches better with Figure\ 1 of \citeA{edberg_statistical_2008}, for which the Mach cone aperture would instead be $\varrho = 13.9$.} \\
        \multicolumn{9}{l}{$^{g}$Fits were performed on 2D-gridded density data, co-depending on $M_{\rm ms}$ and EUV flux levels on the one hand, and solar wind dynamic pressure and EUV flux levels}\\
        \multicolumn{9}{l}{\hspace{2ex} on the other.}\\
        \multicolumn{9}{l}{\hspace{2ex} The coordinate system adopted by \citeA{halekas_structure_2017} was the Mars Solar Electric (MSE) system, with the $X$ axis lying anti-parallel to the solar wind flow.}\\
        \multicolumn{9}{l}{$^{h}$\citeA{ramstad_solar_2017} use the following rectangular function (required to be cylindrically symmetric with respect to the solar wind direction): }\\
        \multicolumn{9}{l}{\hspace{2ex} $\rho = \sqrt{\epsilon^2 - 1}\ \sqrt{\left(x-R_{\rm ss}-\varsigma\right)^2 - \varsigma^2}$, with $\rho$ the radial distance to the bow shock on the $Y^\prime$ axis from the centre of Mars, $R_{\rm ss}$ the subsolar standoff bow shock distance on}\\
        \multicolumn{9}{l}{\hspace{2ex} the $X^\prime$ axis, and $\varsigma$ the so-called scale length. This function is valid $\forall x \neq R_{\rm ss}$ since $\rho(y=0) = R_{\rm ss}$. By definition, $R_{\rm td} = \rho(x=0) = \sqrt{\epsilon^2 - 1}\ \sqrt{\left(R_{\rm ss}+\varsigma\right)^2 - \varsigma^2}$. }\\
        \multicolumn{9}{l}{\hspace{2ex} $\varsigma$ is a constant equal to $33.54\,R_\textrm{p}$ derived in \citeA{ramstad_solar_2017} from the bow shock model values for $R_{\rm ss}$, $R_{\rm td}$ and $\epsilon$ of \citeA{vignes_solar_2000} and can be calculated as}\\
        \multicolumn{9}{l}{\hspace{2ex} $\varsigma = -\frac{1}{2R_{\rm ss}} \left(R_{\rm ss}^2 - R_{\rm td}^2/(\epsilon^2-1)\right)$. The original values of $R_{\rm ss}$ and $\epsilon$ in their study were fitted to a function $a\, n_\text{sw}^b\left(\varv_\text{sw}/100\right)^c + d$; we have assumed here nominal conditions}\\
        \multicolumn{9}{l}{\hspace{2ex} $(n_\text{sw},\varv_\text{sw}) = (2\,\text{cm}^{-3},400\,\text{km\,s}^{-1})$ for simplicity. We calculate the semilatus rectum as $L = \left(R_{\rm td}^2-(\epsilon^2-1)R_{\rm ss}^2\right)/(2R_{\rm ss})$ from formulae\,(\ref{eq:standoffDistanceConic}) and (\ref{eq:standoffTerminatorDistanceConic}). Uncertainties on $\epsilon$}\\
        \multicolumn{9}{l}{\hspace{2ex} fitted values assumed to be $1\%$.}\\
        \multicolumn{9}{l}{$^{i}$Here we only recall the results for all MYs (MY$27-33$). Individual MYs have eccentricities below $1$ (ellipse), except for MY$28-29$ (hyperbola).}\\
        \multicolumn{9}{l}{$^{j}$ That is, Mariner 4 and Mars 2-3-5 \cite{slavin_solar_1991}, Phobos 2 \cite{trotignon_position_1993}, MGS \cite{edberg_statistical_2008}, MEX \cite{hall_martian_2019} and}\\ 
        \multicolumn{9}{l}{\hspace{2ex} MAVEN \cite{halekas_structure_2017}. The listed uncertainties are the standard deviations of the series. Accordingly mean angles $\varrho$ are calculated only for 3 values and}\\
        \multicolumn{9}{l}{\hspace{2ex} are only given for for completeness here.}\\
        \multicolumn{9}{l}{$^{k}$Although this is a hyperbola with cone angle $\varrho = 8.1\deg$, the large eccentricity uncertainty leads to a cone angle uncertainty of $44\deg$, hence no $\varrho$ value is provided here.}\\
        \multicolumn{9}{l}{$^{l}$Calculated from formula\,(\ref{eq:standoffDistanceConic}).}\\
        \multicolumn{9}{l}{$^{m}$Calculated from formula\,(\ref{eq:standoffTerminatorDistanceConic}).}
    \end{tabular}
\end{table*}

\subsubsection{3D Cartesian form}

The more general way of characterising the bow shock shape does not assume any symmetry with respect to any axis. A 3D shape model can be constructed in the form of a quadratic equation \cite<for example,>[for Earth, Mars and comets]{formisano_three-dimensional_1979,gruesbeck_three-dimensional_2018,simon_wedlund_hybrid_2017}:
\begin{linenomath}
\begin{align}
     A x^2 + B y^2 + C z^2 + D xy + E yz + F xz + G x + Hy + Iz - 1 = 0. \label{eq:quadraticBS}
\end{align}
\end{linenomath}
Here and for clarity in the equations, $(x,y,z)$ coordinates are by definition the unaberrated $(X_\textrm{MSO},Y_\textrm{MSO},Z_\textrm{MSO})$ coordinates.
With the MAVEN spacecraft including both magnetometer and ion spectrometer, \citeA{gruesbeck_three-dimensional_2018} used a limited subset of bow shock crossings when ignoring rapid spatial motions of the boundary across the spacecraft due to the variable solar wind \cite<see>{halekas_structure_2017}, leaving a database of only $1799$ crossings spanning about $3$ years of data (November 2014 to April 2017).
For all bow shock detections considered in their study, the best least-squares ellipsoid fit was obtained with $A = 0.049$, $B = 0.157$, $C = 0.153$, $D = 0.026$, $E = 0.012$, $F = 0.051$, $G = 0.566$, $H = -0.031$, $I = 0.019$ and is valid only for the dayside bow shock up to a few $0.1\,R_\textrm{p}$ downstream of the terminator ($X \gtrsim -0.5\,R_\textrm{p}$) because of the poor MAVEN orbital coverage on the nightside flanks of the shock.
\citeA{gruesbeck_three-dimensional_2018} concluded that asymmetry of the shock surface was particularly pronounced in the North-South direction, likely due to the influence of crustal magnetic fields. Located predominantly in the southern hemisphere of Mars between $30\deg$S--$85\deg$S \cite{Acuna1998}, they tend to increase the altitude of the induced magnetospheric boundary and hence increase the subsolar standoff distance of the shock \cite<see>[and references therein]{Matsunaga2017}. 

Although quadratic surfaces are not necessarily centered on the planet nor is their main axis directed along the $X_{\rm MSO}$ axis (see \ref{sec:appendix1}), a simple estimator of the shock's position in the subsolar and terminator directions can be of interest. We propose here such an expression, based on Equation\,(\ref{eq:quadraticBS}). We calculate thus the subsolar standoff distance along the $X_{\rm MSO}$ axis at coordinates ($x$, $y=0$, $z=0$) by finding the positive root of the simplified quadratic equation (i.e. the intersection of the surface with the $X_{\rm MSO}$ axis):
\begin{linenomath}
\begin{align}
    A x^2 + G x - 1 = 0\\
    \Longrightarrow x_\textrm{max} = R_\textrm{ss} = \frac{-G + \sqrt{G^2+ 4A}}{2A}, \label{eq:Rss}
\end{align}{}
\end{linenomath}
whereas the terminator standoff distances in the $Y-Z$ plane (non-aberrated MSO coordinates) are similarly given by:
\begin{linenomath}
\begin{align} 
    B y^2 + H y - 1 = 0\quad \Longrightarrow&\quad y_\textrm{max} = R_{\textrm{td},y} = \frac{-H + \sqrt{H^2+ 4B}}{2B}\label{eq:Rtdy}\\
    C z^2 + I z - 1 = 0\quad \Longrightarrow&\quad z_\textrm{max} = R_{\textrm{td},z} = \frac{-I + \sqrt{I^2+ 4C}}{2C} \label{eq:Rtdz}
\end{align}{}
\end{linenomath}
Because of the small angles involved at Mars, non-aberrated coordinates are rather accurate for the subsolar standoff distance. That said, rotating the MSO coordinate system by a certain small angle $\alpha$ about the $Z$ axis does impact the terminator distances by a few $0.01\,R_\textrm{p}$. For the parameters given above, $R_\textrm{ss} = 1.56\,R_\textrm{p}$,  $R_{\textrm{td},z} = 2.50\,R_\textrm{p}$ and $R_{\textrm{td},y} = 2.62\,R_\textrm{p}$. Because the shock is a 3D object, the exact position of the tip of the ellipsoid may vary with respect to the values taken at the origin.

\subsubsection{Comparison of historical models}

\begin{figure}[htb]
 \includegraphics[width=\columnwidth]{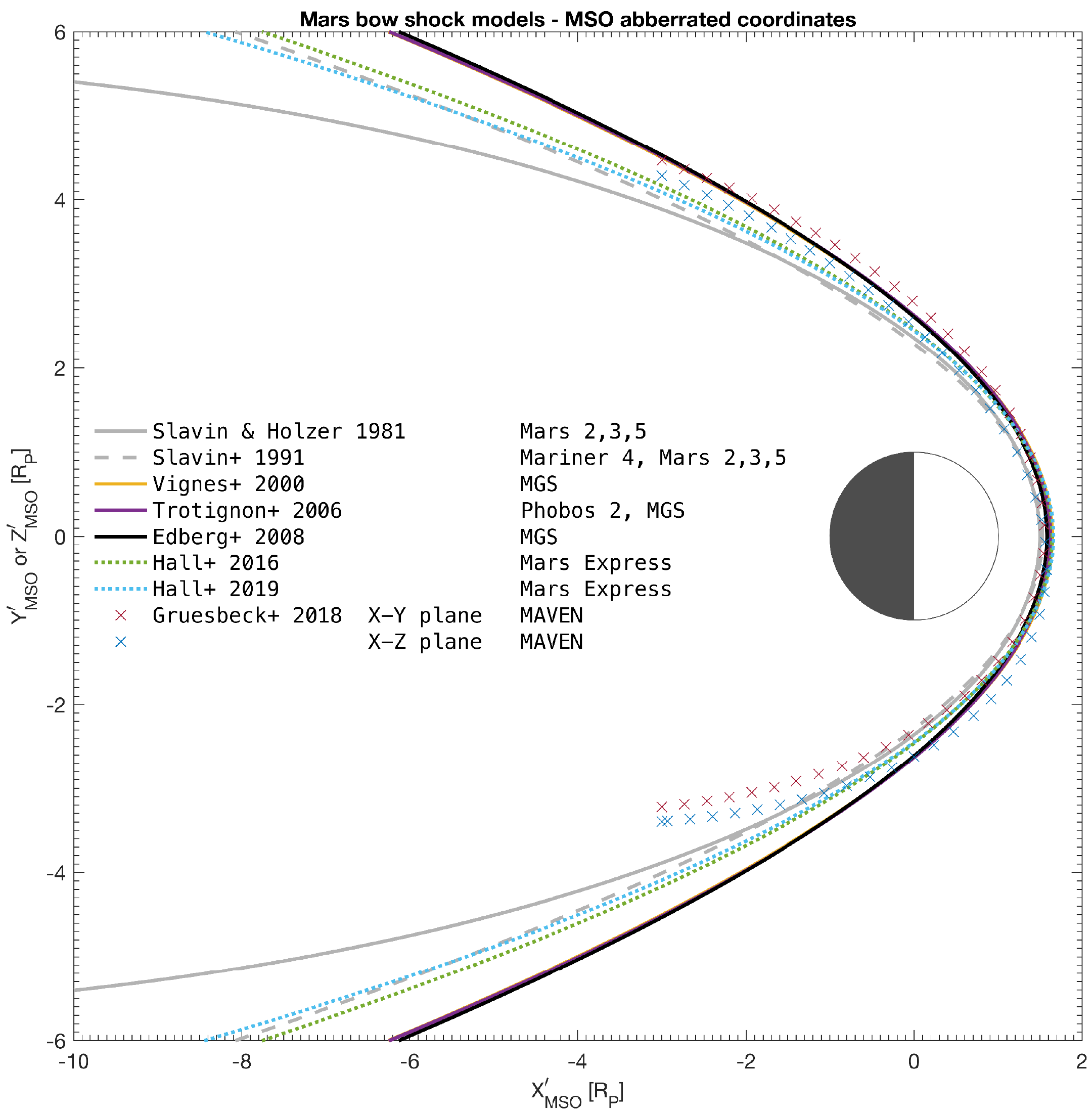}
 \caption{Bow shock fitted models to observations in MSO aberrated coordinates. The 3D quadratic model of \citeA{gruesbeck_three-dimensional_2018} fitted from MAVEN data was rotated anticlockwise $4\deg$ around the $Z$ axis. All other models are obtained in cylindrical conic form from other missions, including Mars Express (MEX), Mars Global Surveyor (MGS), Phobos 2 and the Mars 2-3-5 missions. The fits of \citeA{hall_annual_2016}, \citeA{hall_martian_2019} and \citeA{gruesbeck_three-dimensional_2018} consider all shock detection points of their respective studies. Because the cylindrical models are symmetric about the $X^\prime$ axis, the figure's cylindrical $y$-axis $\sqrt{Y^{\prime2}+Z^{\prime2}}$ is equivalent to the $Y^\prime$ or to the $Z^\prime$ axis, regardless. The coordinates are normalised to the radius of Mars, $R_\textrm{p} = 3,389.5$\,km. \label{fig:bowShockModels}
 }
\end{figure}

A comparison of a representative selection of historical bow shock models \cite<some of them as listed in>[]{vignes_solar_2000,trotignon_martian_2006} in the $X^\prime-Y^\prime/Z^\prime$ plane is given in Figure\ \ref{fig:bowShockModels}. The 3D quadratic model of \citeA{gruesbeck_three-dimensional_2018} was rotated anticlockwise by $4\deg$ around the $Z$ axis to ease the comparison. It is noteworthy to remark that although the fit is not valid for $X^\prime\lesssim-0.5\,R_\textrm{p}$, the figure displays the fits for $X^\prime>-2.5\,R_\textrm{p}$ to illustrate the differences in shock surface swelling.

All models are in excellent agreement around the subsolar point, with a mean subsolar standoff distance value of $R_\mathrm{ss}=1.59\pm0.05\,R_\textrm{p}$. The terminator standoff distance is also in very good agreement -- however, for $X^\prime\lesssim0\,R_\textrm{p}$, the difference between fits becomes substantial, especially $(i)$ between the recent MEX investigations of \citeA{hall_annual_2016,hall_martian_2019} and all other fits on the one hand, and $(ii)$ between the MAVEN fits and the other fits on the other. For MAVEN, this is due, as previously mentioned, to the lack of orbital coverage by the spacecraft for $X^\prime< 0\,R_\textrm{p}$. In this sense, MEX has a much better antisolar spatial coverage.
Of note, the North-South asymmetry in the 3D fits of \citeA{gruesbeck_three-dimensional_2018}, coinciding with the presence of crustal fields in the Southern hemisphere of Mars, can easily be seen, a characteristic which no axisymmetric model directly quantifies.

\subsection{Shock geometry}\label{sec:geometry}

\subsubsection{Quasi-perpendicular or quasi-parallel shock?}
A collisionless shock may have different behaviours depending on the upstream solar wind magnetic field (the IMF), which conditions how the solar wind is losing its energy to the magnetosheath. Two main cases are conveniently studied for their varying properties: $q_\parallel$ and $q_\perp$ shocks. Additional important physical quantities driving the shock structure and dynamics are the magnetosonic Mach number (which defines the shock's criticality) and the plasma-$\beta$ \cite{balogh_physics_2013}. 

It is useful to recall that a $q_\parallel$ shock condition is defined so that the background IMF lines are intersecting normally the shock surface, whereas a $q_\perp$ shock describes an IMF that is in effect {\it in the tangent plane} to the surface shock. 
Thus, the angle of importance is the angle between the average IMF vector upstream of the shock and the shock normal. This angle is in the literature almost always named $\theta_{Bn}$, which is kept here for convenience. The geometry of the shock is defined as follows:
\begin{linenomath}
\begin{align}
	\theta_{Bn} &> 45\deg: \textrm{$q_\perp$ shock}\\
	\theta_{Bn} &\leq 45\deg: \textrm{$q_\parallel$ shock}
\end{align}
\end{linenomath}
Starting in the magnetic field compression region in the solar wind, $q_\perp$ shocks have structures, from the point of view of $B$-fields, almost always characterised by $(i)$ a foot, $(ii)$ a fast ramp, and $(iii)$ a wider overshoot followed by a more gradual undershoot \cite<see>[Figure\ 11]{kennel_quarter_1985}. This classic picture is a first approximation as fine electron-scale structures in the foreshock, foot and ramp can be seen with high-cadence magnetic field measurements. $Q_\perp$ shocks reflect particles back upstream to satisfy the shock conditions and are on average diffusive. Magnetic structures trapping particles such as mirror modes are observed to predominantly take place in the magnetosheath behind a $q_\perp$ shock \cite{gary_mirror_1992}.
On the other hand, $q_\parallel$ shocks are on average resistive and are usually characterised by heavy turbulence. Their foreshock contains MHD turbulence that can give rise to first-order Fermi acceleration. Also common in the foreshock region, highly compressive structures such as Short Large-Amplitude Magnetic Structures (SLAMS) are associated with large density variations: they originate from the steepening of ULF waves and are of great importance in the shock reformation \cite{burgess_quasi-parallel_2005,burgess_microphysics_2014}.

In order to determine the $q_\parallel$ or $q_\perp$ geometry of the shock crossed by a spacecraft, the normal direction to the shock surface needs to be first estimated. For a single spacecraft, this can be done either with methods that take advantage of upstream and downstream magnetic field measurements \cite<coplanarity method as in>[although prone to rather large uncertainties]{schwartz_shock_1998,horbury_four_2002} or through geometrical considerations only, as we propose in Section\ \ref{sec:shocknormal}. 
The accuracy of the geometrical method presented here is linked to the assumption that the shock surface is smooth and does not possess any kinks or local structures where the current curls on itself. In practice, we do not expect such a smooth surface as the shock may assume a more rippled shape which depends on the upstream solar wind condition and the turbulence at the boundary \cite{moullard_ripples_2006}. However, our geometric determination may still be a useful first approximation of the geometry of the shock.

\subsubsection{Determination of the shock normal}\label{sec:shocknormal}

The normal to the shock surface at point $(r_0,\theta_0)$ in polar coordinates, $(x_0,y_0,z_0)$ in Cartesian coordinates or $(r_0,\vartheta_0,\varphi_0)$ in spherical coordinates, is simply defined as the gradient vector of the (assumed) smooth surface $f$ at that point. Mathematically we can express this condition as:
\begin{linenomath}
\begin{align}
    \nabla f \cdot \v{v} = 0
\end{align}
\end{linenomath}
where $\v{v}$ is a vector tangential to the surface at that point. This leads to the following expressions in the 2D and 3D cases.

{\bf 2D case}. For the 2D polar coordinate case, let $f$ be equal to $f(r,\theta) = r - L/(1 + \epsilon\cos\theta)$ following Equation\,(\ref{eq:polarBSmodel}) where $\theta$ is the angle from the focus $x_F$ on the $X^\prime$ axis. The gradient of $f$ depends on the two variables $(r,\theta)$:
\begin{linenomath}
\begin{align}
    \nabla f = \begin{pmatrix}
        \pd{f}{r}\\
        \frac{1}{r}\pd{f}{\theta}
        \end{pmatrix}
        =
        \begin{pmatrix}
            1\\
            -\frac{\epsilon \sin\theta}{\left(1+\epsilon\cos\theta\right)}
         \end{pmatrix}\label{eq:normalPolar2D}
\end{align}
\end{linenomath}

At point $(r_0,\theta_0)$ vector $\nabla f = (R_0,\Theta_0)$ is perpendicular to the surface. Note that because of the peculiarity of a conic, values in $x$ must always be corrected by the focus distance $x_F$, because the typical polar angle $\vartheta$ is not strictly the same as the conic angle $\theta$ used in Equation\,(\ref{eq:polarBSmodel}) (see Figure\ \ref{fig:2dconic}).

\begin{figure*}[tb]
 \includegraphics[width=\textwidth]{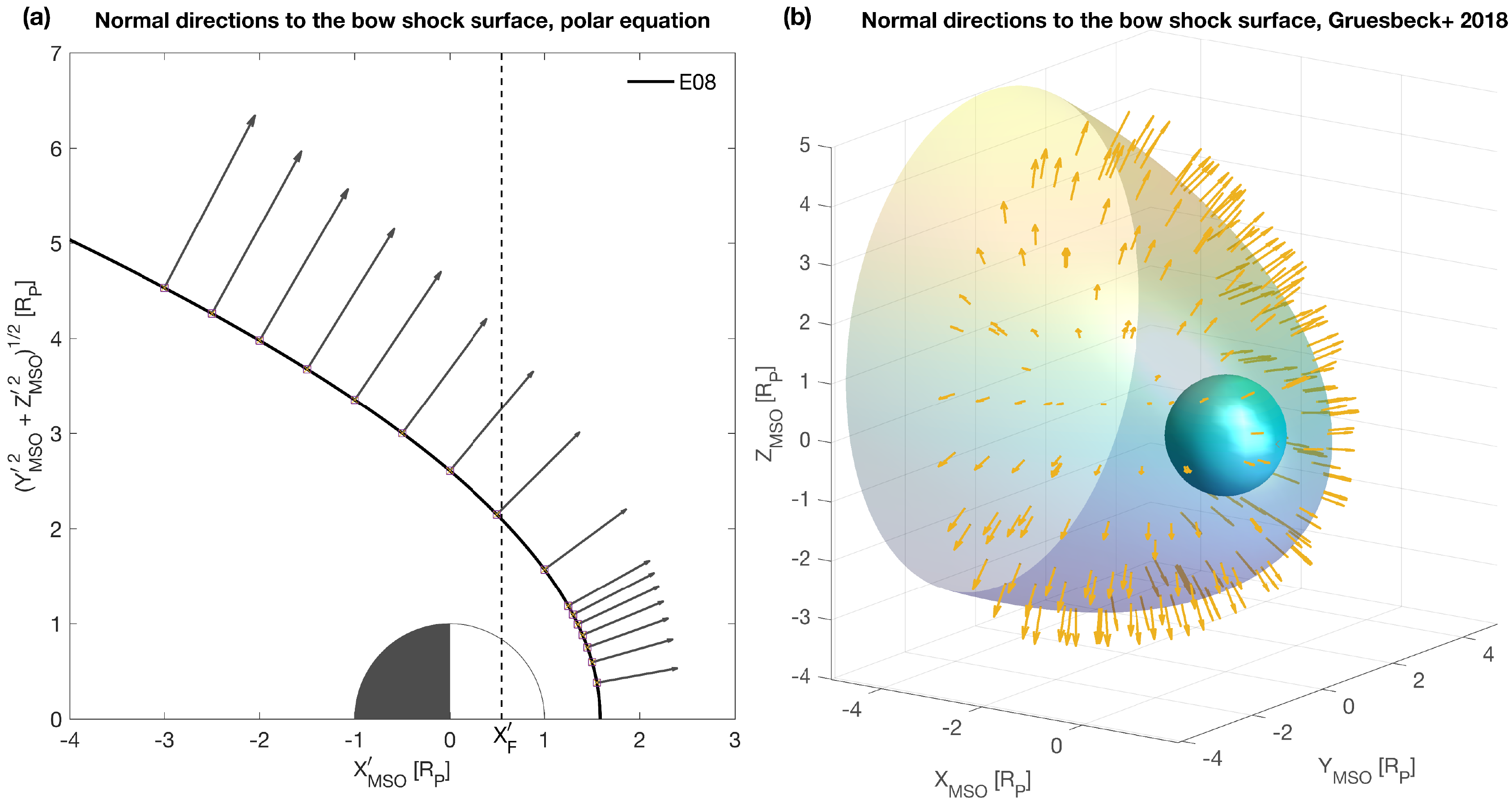}
 \caption{Bow shock normal. (a) Polar 2D case \cite<model of>[]{edberg_statistical_2008}. (b) Cartesian 3D case \cite<model of>[]{gruesbeck_three-dimensional_2018}. All coordinates are normalized to the radius of Mars, $R_\textrm{p} = 3,389.5$\,km.
 \label{fig:normalBowShock}
 }
\end{figure*}

In Cartesian coordinates, using Equation\,(\ref{eq:cartBSmodel2D}), the gradient will be against directions along $X^\prime$ and $Y^\prime$ and equal to:
\begin{linenomath}
\begin{align}
    \nabla f = \begin{pmatrix}
        \pd{f}{X^\prime}\\
        \pd{f}{Y^\prime}
        \end{pmatrix}
        =
        \begin{pmatrix}
           -2\left(\epsilon^2 - 1\right)\,\left(X^\prime - x_F\right) + 2\epsilon L \\
            2 Y^\prime
         \end{pmatrix}.\label{eq:normalCartesian2D}
\end{align}
\end{linenomath}
This expression circumvents the ambiguity on the angle direction of the polar formula, and as such should be preferred when calculating the normal direction. Figure\ \ref{fig:normalBowShock}a displays our estimates of the normal direction to several points on the shock surface applied to the 2D bow shock polar fit of \citeA{edberg_statistical_2008}, and converted to Cartesian coordinates.

{\bf 3D case}. For the 3D Cartesian quadratic equation, $f$ is simply equal to the left member of equation\,(\ref{eq:quadraticBS}).

The gradient of $f$ is then:
\begin{linenomath}
\begin{align}
    \nabla f = \begin{pmatrix}
        \pd{f}{x}\\
        \pd{f}{y}\\
        \pd{f}{z}
        \end{pmatrix}
        =
        \begin{pmatrix}
           2Ax + Dy  +  Fz + G \\
            Dx + 2By +  Ez + H \\
            Fx + Ey  + 2Cz + I
         \end{pmatrix}\label{eq:normalCartesian3D}
\end{align}
\end{linenomath}
At point $(x_0,y_0,z_0)$ vector $\nabla f = (X_0,Y_0,Z_0)$ is perpendicular to the surface. Additionally, the equation of the tangent plane to the smooth surface at that point is of the general form:
\begin{linenomath}
\begin{align}
    \left.\pd{f}{x}\right|_{0} \left(x-x_0\right) + \left.\pd{f}{y}\right|_{0} \left(y-y_0\right) + \left.\pd{f}{z}\right|_{0} \left(z-z_0\right) = 0,
\end{align}
\end{linenomath}
where subscript "0" in the gradient components denotes the gradient taken at points $(x_0,y_0,z_0)$ for brevity.
Thus, knowing the 3D position of the spacecraft at the expected bow shock position, one can calculate the transverse and tangent directions to the bow shock surface.
Figure\ \ref{fig:normalBowShock}b shows our results with this technique applied to the 3D bow shock fit of \citeA{gruesbeck_three-dimensional_2018}, assuming a spacecraft situated on random points of the shock's surface.

With the knowledge of the normal direction to the shock at the spacecraft location, it becomes possible, from the 2D and 3D model cases, to calculate the angle $\theta_{Bn}$ from the average direction of the magnetic field at the spacecraft position near the shock, noted $\mathbf{B}_\text{bg}$: 
\begin{linenomath}
\begin{align}
	\tan \theta_{Bn} =  \frac{||\nabla f \times \mathbf{B}_\text{bg}||}{\nabla f \cdot \mathbf{B}_\text{bg}}.
\end{align}
\end{linenomath}
For added robustness, the function $\arctantwo$ is recommended for the calculation of the inverse tangent, as it returns a value corresponding to the correct quadrant of the Euclidean plane.
Because of the inherent 3D nature of a spacecraft orbit and of the local magnetic field, we prefer the 3D calculation with Equation\,(\ref{eq:normalCartesian3D}) over the corresponding 2D case (Equation\,\ref{eq:normalCartesian2D}). It is however important to recall here that the shock's local shape may be assuming that of a ``corrugated iron'' section, as evidenced for example at Earth with the Cluster quartet of spacecraft \cite{moullard_ripples_2006}. No method is foolproof in estimating $\theta_{Bn}$: in case studies, the local normal to the shock can be more carefully checked, for which several complementary methods exist, such as the magnetic coplanarity method mentioned earlier. 

When applied to the MAVEN dataset, the geometric calculation of $\theta_{Bn}$, assuming a smooth surface, is expected to reach an uncertainty of about $\pm 5\deg$ depending on the upstream field determination. We obtained this estimate for a small sample of crossings, by extending over a few minutes the time spans used to calculate the upstream magnetic field direction.

\section{Detecting the bow shock in spacecraft orbits}\label{sec:bowshockDetection}
Estimating the bow shock position from spacecraft spatial coordinates can be achieved either empirically or theoretically, depending on the precision needed. Semi-empirical but computationally intensive techniques using machine-learning imaging algorithms are currently attempted to detect automatically and precisely the exact position of the shock. Such technique makes use of the full plasma instrumental payload on board planetary missions, when available. Other techniques like that of \citeA{Nemec2020} use ion and magnetic field data in combination to statistically identify the regions crossed by the spacecraft. However, a faster approach, based on a simple geometrical estimator using a static analytical bow shock model (see Section\ \ref{sec:bowshockModels}), may still prove valuable for statistical studies or for new datasets. We present such an approach and its possible refinements in 2D and 3D coordinate systems based on magnetic field-only measurements.
Because in the following we do not take into account plasma measurements, and because the true signature of the shock may be difficult to detect with magnetic field data only, additional criteria to identify solar wind and magnetosheath regions are required to mitigate this ambiguity. These criteria are presented in Section\ \ref{sec:PredictorCorrector}.

\subsection{Predictor algorithm for the shock position from existing analytical models}\label{sec:Predictor}

The predictor algorithm is based on the calculation of polar (2D, $\theta$) or spherical (3D, [$\vartheta$, $\varphi$]) angles in the corresponding frame of reference centred on the planet in MSO coordinates. These angles unequivocally define the predicted distance to the bow shock at the position of the spacecraft. Comparing this bow shock distance with the Euclidean spacecraft distance to the centre of the planet gives access to the region in which the spacecraft is located either in the solar wind or in the magnetosheath. 

{\bf 2D case}. 
Our algorithm is (see Figure\ \ref{fig:2dconic} for definitions of angle and distances):
\begin{itemize}
    \item Choose conic model $r(\theta)$, with eccentricity $\epsilon$, semilatus rectum $L$ and focus' position ($x_F, 0, 0$),
    \item Calculate the spacecraft's Euclidean distance $r_\textrm{sc}$ from the chosen conic model focus $x_F$, in aberrated MSO coordinates, so that: $r_\textrm{sc}=\sqrt{ \left(X_\textrm{sc}^{\prime}-x_F\right)^2+Y_\textrm{sc}^{\prime 2}+Z_\textrm{sc}^{\prime 2} }$, 
    \item Calculate the angle $\theta_\textrm{sc}$ at the position of the spacecraft: $\theta_\text{sc}= \arctantwo\left(\sqrt{Y_\textrm{sc}^{\prime 2}+Z_\textrm{sc}^{\prime 2}},\left(X_\textrm{sc}^\prime-x_F\right)\right)$,
    \item Calculate the predicted bow shock distance $R_\textrm{bs}$ at the corresponding spacecraft $\theta_\textrm{sc}$ from the focus $x_F$: $R_\textrm{bs} = r(\theta_\textrm{sc}) = L / (1 + \epsilon \cos{\theta_\text{sc}} )$ following Equation\,(\ref{eq:polarBSmodel}),
    \item Calculate $\Delta R = r_\textrm{sc} - R_\textrm{bs}$. If $\Delta R$ goes from negative to positive (respectively, from positive to negative) values, the spacecraft is expected to move from the magnetosheath to the solar wind (respectively, from solar wind to magnetosheath). At the temporal precision of the spacecraft ephemerides, the closest value to $\Delta R=0$ defines the shock position $\left(X_\textrm{bs}, Y_\textrm{bs}, Z_\textrm{bs}\right)$ and crossing time $t_\textrm{bs}$.
\end{itemize}

This purely geometrical approach was tested for polar coordinate models \cite<such as>[]{edberg_statistical_2008,hall_annual_2016,hall_martian_2019}, provided that all spacecraft coordinates are first rotated $4\deg$ into the aberrated MSO system. 

{\bf 3D case}.
For 3D quadric models \cite{gruesbeck_three-dimensional_2018}, the aberration is already taken into account and there is no need to correct the spacecraft coordinates for the position of the focus of the conic. Thus we only need generalise the approach above to spherical coordinates ($\rho$, $\vartheta$, $\varphi$), where $\rho$\,$=$\,$\sqrt{X^2+Y^2+Z^2}$ is the planetocentric distance, whereas $\vartheta$\,$=$\,$\arctan{\left(Y/X\right)}$ and $\varphi$\,$=$\,$\arctan{\left(Z/\sqrt{X^2+Y^2}\right)}$ represent azimuth and elevation (measured from the $X$--$Y$ plane) by convention. To compensate for the inherent ambiguity on azimuth depending on the quadrant and gain robustness, the function $\arctantwo$ is preferred throughout for simplicity, in a programming sense. 
Equation\,(\ref{eq:quadraticBS}) becomes a second-degree equation of the form: 
\begin{linenomath}
\begin{align}
&  a \rho^2 + b \rho - 1 = 0\label{eq:quadricSpherical}
\end{align}
\end{linenomath}
with:
\begin{linenomath}
\begin{align*}
& a = A\cos^2{\varphi}\,\cos^2{\vartheta} + B\cos^2{\varphi}\,\sin^2{\vartheta} + C\sin^2{\varphi}\\
&    + \frac{D}{2}\cos^2{\varphi}\,\sin{2\vartheta} + \frac{E}{2}\,\sin{2\varphi}\,\sin{\vartheta} + \frac{F}{2}\,\sin{2\varphi}\,\cos{\vartheta}\\
\textrm{and:}\\
& b = G \cos{\varphi}\,\cos{\vartheta} + H\cos{\varphi}\,\sin{\vartheta} + I\sin{\varphi}
\end{align*}{}
\end{linenomath}
In the case of an ellipsoid of revolution \cite<as it is the case for the parametrisation of>[]{gruesbeck_three-dimensional_2018}, the bow shock distance $R_\textrm{bs}$ at the angles ($\vartheta$,$\varphi$) corresponds to the positive root of this equation:
\begin{linenomath}
\begin{align}
    R_\textrm{bs} = \frac{-b+\sqrt{4 a + b^2}}{2a}, \label{eq:standoffBS}
\end{align}{}
\end{linenomath}
for $a\neq 0$. For other parametrisations such as a hyperboloid of two sheets, there may be two positive roots, in which case the smallest root should be chosen.

The denominator $2a$ in expression\,(\ref{eq:standoffBS}) never reaches zero, no matter the combination of angles chosen, which makes it a robust formula throughout any orbit. For azimuth and elevation angles ($|\vartheta| \gtrsim115\deg, |\varphi|\lesssim55\deg$), the model of \citeA{gruesbeck_three-dimensional_2018} is not applicable any more (standoff distances above $4\,R_\textrm{p}$) as these particular angular combinations correspond to a tail-flank position, lying outside of MAVEN's orbital range.

The next step is to determine whether the spacecraft is inside the bow shock surface or outside of it in the orbital sequence. 
Our algorithm follows a similar sequence as for the 2D case, but all variables are calculated with respect to the centre of Mars, in unaberrated MSO coordinates:
\begin{itemize}
    \item Choose 3D model of the bow shock with $A, B, C, D, E, F, G, H$ and $I$ parameters,
    \item Calculate the spacecraft's Euclidean distance in non-aberrated coordinates, $R_\textrm{sc} = \sqrt{X_\textrm{sc}^{2}+Y_\textrm{sc}^{2}+Z_\textrm{sc}^{2}}$, 
    \item Calculate (azimuth, elevation) angles ($\vartheta, \varphi$) at position of the spacecraft: $\vartheta_\text{sc} = \arctantwo\left(Y_\textrm{sc},X_\textrm{sc}\right)$ and $\varphi_\text{sc} = \arctantwo\left(Z_\textrm{sc},\sqrt{X_\textrm{sc}^2+Y_\textrm{sc}^2}\right)$,
    \item Calculate the bow shock distance $R_\textrm{bs}$ at the corresponding spacecraft spherical angles with Equation\,(\ref{eq:standoffBS}),
    \item Calculate $\Delta R = R_\textrm{sc} - R_\textrm{bs}$.  If $\Delta R$ goes from negative to positive (respectively, from positive to negative) values, the spacecraft is expected to move from the magnetosheath to the solar wind (respectively, from solar wind to magnetosheath). At the temporal precision of the spacecraft ephemerides, the closest value to $\Delta R=0$ defines the shock position $\left(X_\textrm{bs}, Y_\textrm{bs}, Z_\textrm{bs}\right)$ and crossing time $t_\textrm{bs}$.
\end{itemize}

\begin{figure}[tb]
 \includegraphics[width=\columnwidth]{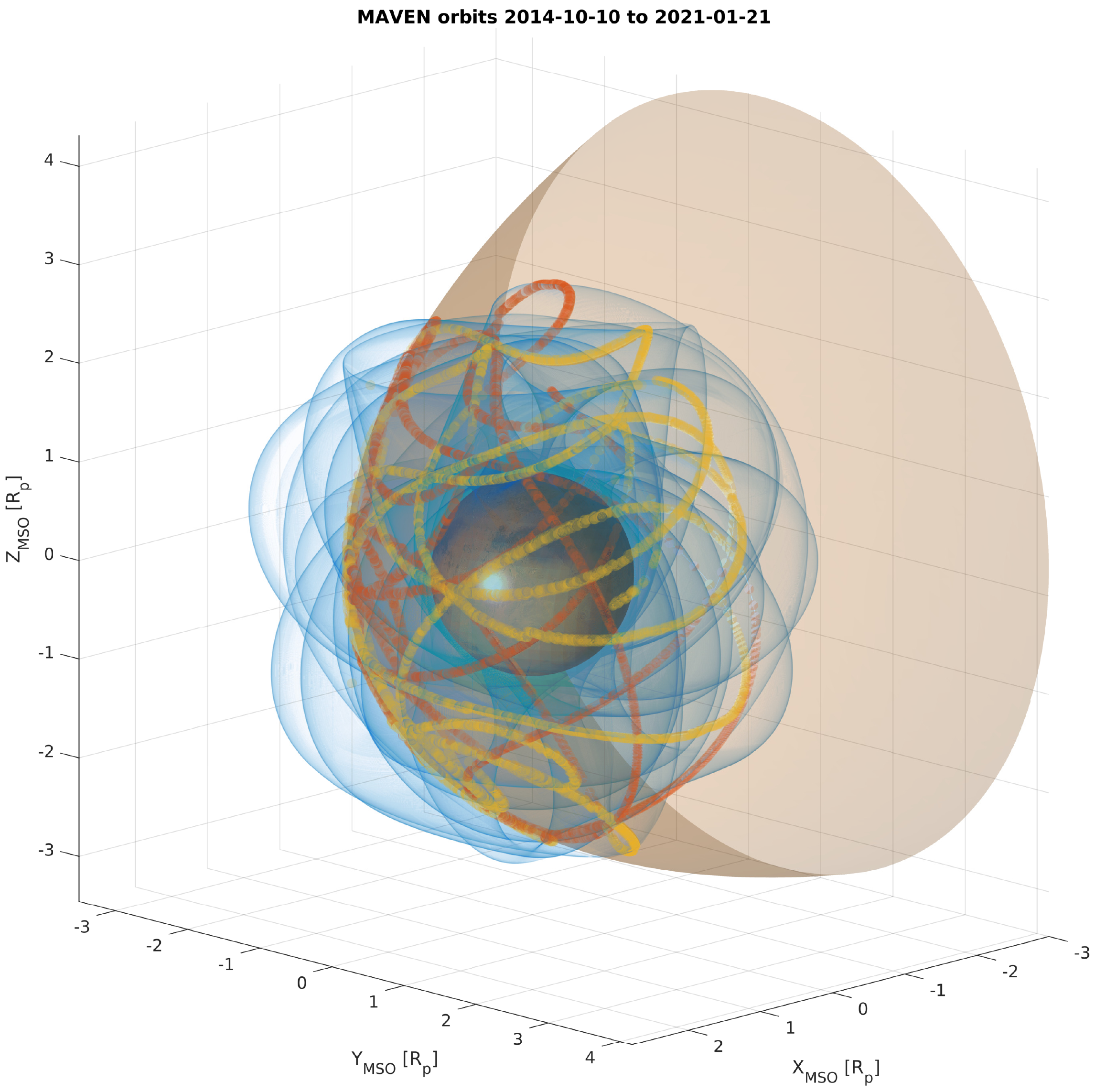}
 \caption{Automatic detection of bow shock in MSO coordinates normalised to the planet's radius, using the general quadric formula of \citeA{gruesbeck_three-dimensional_2018}. The bow shock surface is in brown, the orbit of MAVEN between 01 November 2014 and 07 February 2021 is in blue. Detections of the crossings from inside the shock surface to outside of it are shown as orange circles, whereas outside-to-inside crossings are depicted by yellow circles. Coordinates are normalised to the radius of Mars, $R_\textrm{p} = 3,389.5$\,km.
  \label{fig:BSdetect}
 }
\end{figure}

{\bf Application to MAVEN orbits.} 
Using 1-min-averaged MAVEN orbits, we perform the automatic detection of the bow shock as shown in Figure\ \ref{fig:BSdetect}. Because of the relatively poor temporal resolution of this dataset, as well as the fast approach in the early stages of the orbit insertion, some points in the orbit yield false positive detections which disappear when increasing the orbital data resolution to \qty{1}{s}.

Thanks to the simple algorithms presented above, we can statistically predict bow shock crossings in a given spacecraft orbit. To help identify the solar wind region, we can distinguish between trajectories moving from the magnetosheath to the solar wind region, and vice-versa. For each orbit intersecting the bow shock model, two points per orbit will be identified. At \qty{1}{s} resolution, we predicted a total of $16,515$ bow shock crossings using the 3D analytical model of \citeA{gruesbeck_three-dimensional_2018} for the MAVEN dataset between November 2014 and February 2021, including $8,256$ crossings from the sheath to the solar wind and $8,259$ crossings from the solar wind to the sheath.

Actual crossings may in practice be different than the predictions and the algorithms may fail to pinpoint the location of the shock sometimes by several tens of minutes, mostly because of solar wind varying conditions, or due to the geometry of the shock at the point of passage for an individual orbit \cite{halekas_structure_2017}. We estimate thus the precision of these automatic estimates to be of the order of \qty{\pm 0.08}{$R_p$} (\qty{\pm 270}{km}) around the `true' bow shock location, from a representative subset of data. Because of variable shock position from orbit to orbit and the geometric average nature of the detection, some orbits that may have experienced shock crossings but lie inside the average shock position will not be tested for potential detection. It is estimated that only a few hundred potential crossings were ignored in the process.
Consequently, the true shock structure location should be checked directly in the magnetic field and plasma data. Moreover, since the bow shock is a dynamical object, it may experience fast forward and backward motions, crossing the spacecraft trajectory several times more per orbit. This can be seen for example in Figure\ \ref{fig:BowShockExamples}c, where the total magnetic field undergoes sharp intermittent jumps in the foreshock area (23 June 2018, around 01:00\,UT). As in \citeA{gruesbeck_three-dimensional_2018}, these multiple crossings, usually the first one in the temporal sequence for crossings into the sheath, and the last one for crossings into the solar wind, are identified as one with our algorithm.

We successfully applied this 3D algorithm to the retrieval of undisturbed solar wind density and velocity moments in the MAVEN/SWIA data with \qty{1}{min} resolution, as part of the Helio4Cast solar wind in-situ data catalogue, enabling the statistical study of interplanetary coronal mass ejections and high speed streams \cite{mostl_solar_2020}.

\subsection{Refining the position of the shock: a predictor-corrector algorithm}\label{sec:PredictorCorrector}
Because of variations in the shock position, the automatic detection may give inaccurate predictions. We present here a fast method to correct to a certain extent for these discrepancies. It makes use of the magnitude of $\mathbf{B}$ to identify the position of the shock structure, either when crossing from the magnetosheath into the solar wind or vice-versa. As before, the main assumption is that the shock is crossed twice per orbit at maximum, although in practice the shock structure may be crossed several times due to the fast motion of the boundary across the spacecraft trajectory \cite{halekas_structure_2017}. As we are interested in the statistical position of the shock, this assumption nonetheless provides a valuable estimate of the average position of the shock during those times. All magnetic field data are assumed here to be of the order of $1$-s resolution. 

\begin{figure*}[tb]
 \includegraphics[width=\textwidth]{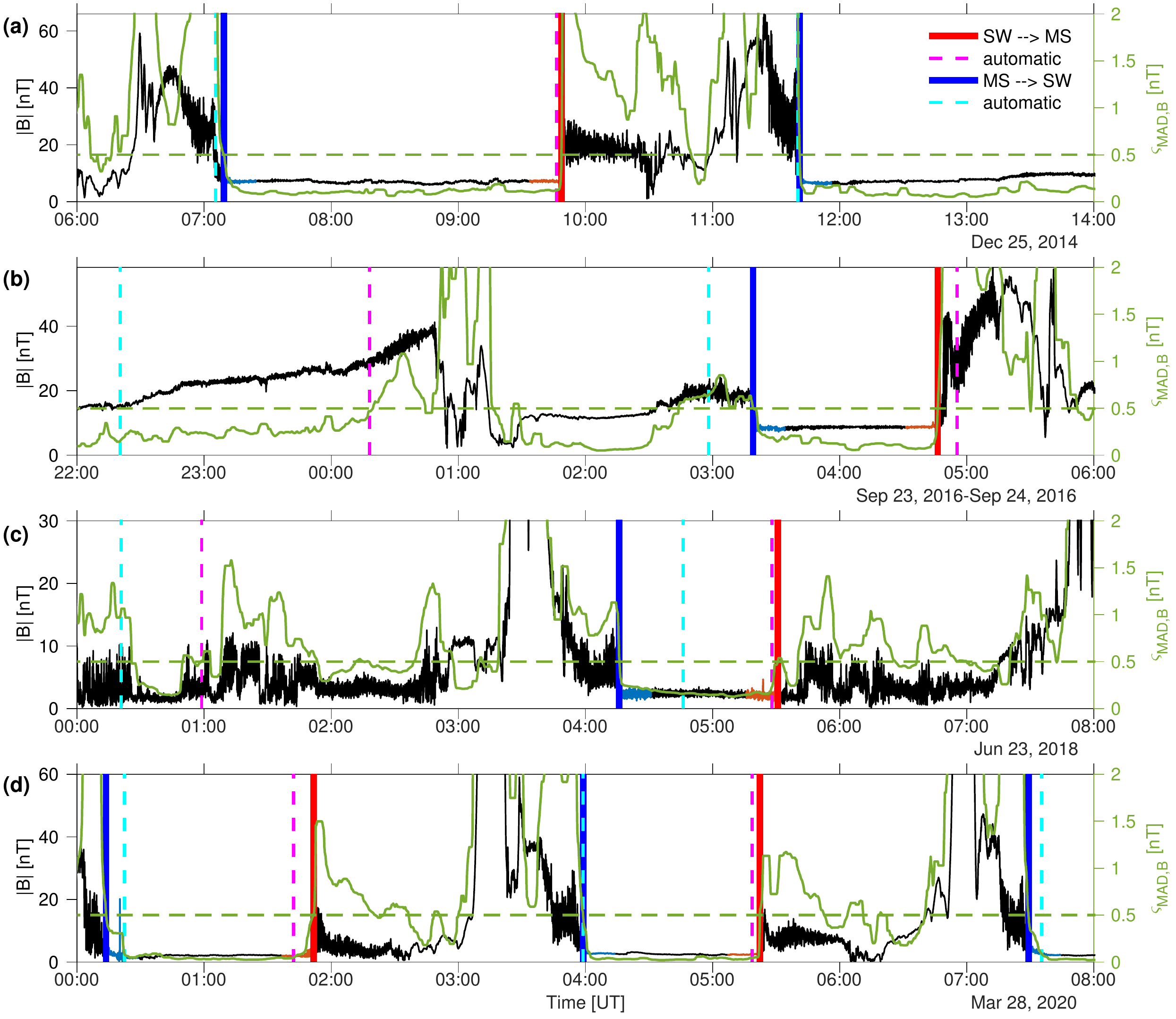}
 \caption{Examples of total magnetic field amplitudes $|\mathbf{B}|$ at $1$\,s resolution measured by the MAVEN/MAG instrument throughout the mission (left $y$-axes), and calculated running median absolute deviations $\varsigma_\textrm{mad,B}$ (right $y$-axes in green). (a) 25 December 2014 (beginning of mission). The first crossing is quite oblique ($\theta_{B{\rm n}}$\,$\approx$\,$45\deg$) followed by two highly $q_\perp$ shock crossings ($\theta_{B{\rm n}}$\,$>$\,$85\deg$). (b) 23--24 September 2016. The two detected crossings are  $q_\perp$, the first one with $\theta_{B{\rm n}}$\,$\approx$\,$58\deg$, the second with $\theta_{B{\rm n}}$\,$\approx$\,$78\deg$. (c) 23 June 2018, with two detected $q_\parallel$ crossings ($\theta_{B{\rm n}}$\,$\approx$\,$8, 25\deg$). (d) 28 March 2020, with five crossings all oblique towards $q_\perp$ conditions, with $\theta_{B{\rm n}}$\,$\approx$\,$45\deg, 49\deg, 80\deg, 82\deg$ and $88\deg$, successively. The predictor geometric detections (Section\ \ref{sec:Predictor}) are in dashed lines and labelled ``automatic'', whereas the predictor-corrector detections proposed in Section\ \ref{sec:PredictorCorrector} are in solid lines. Highlighted in different colours are crossings from solar wind to magnetosheath (labelled SW\,$\rightarrow$\,MS, red) and from magnetosheath to solar wind (labelled MS\,$\rightarrow$\,SW, blue). Calculations of $\theta_{B{\rm n}}$ angles were performed using median averages of $\v B$ over the colour-highlighted regions (blue for MS\,$\rightarrow$\,SW crossings, red for SW\,$\rightarrow$\,MS crossings). The threshold $\varsigma_\textrm{th}$\,$=$\,$0.5$ is shown as a horizontal dashed line (right $y$-axis, green).  
 \label{fig:BowShockExamples}
 }
\end{figure*}

From the point of view of a single spacecraft's magnetic field measurements across the shock boundary, the total magnetic field intensity increases sharply, from typically $5$--$10$ nT in the solar wind at Mars to about twice that level on average when moving into the magnetosheath. Additionally, fluctuations increase, going from small standard deviations around the mean in the solar wind, to comparatively larger fluctuations in the magnetosheath. Typical crossings of the Martian shock illustrating these behaviours are shown for example in Figure\ \ref{fig:BowShockExamples} which presents examples of bow shock crossings around Mars as seen with the MAG instrument on board MAVEN throughout the mission. As discussed in Section\ \ref{sec:geometry}, $q_\perp$ shocks usually display distinct features (foot, sharp ramp, overshoot) as in Figure\ \ref{fig:BowShockExamples}a,b,d), whereas $q_{||}$ geometries have less clear signatures in magnetic field data, making it difficult to detect reliably (Figure\ \ref{fig:BowShockExamples}c). Consequently, the corrector method presented here is biased towards the detection of $q_\perp$ crossings.

Our predictor-corrector detection algorithm attempts to consistently identify solar wind undisturbed regions in the chosen dataset, characterised by relatively low magnetic field intensities in combination with small fluctuations. For each orbit of a spacecraft, it proceeds as follows, with user-defined values ($\Delta t_\textrm{bs}$, $B_\textrm{th}$, $T$, $\varsigma_\textrm{th}$) discussed afterwards:
\begin{enumerate} 
	\item Calculate predictor estimate of the shock crossing time $t_\textrm{bs}$ with the automatic algorithm in 2D aberrated polar coordinates or in 3D (as in Section\ \ref{sec:Predictor}).
	\item If $t_\textrm{bs}$ exists in the considered orbit, choose a small time interval symmetric around the estimated shock so that [$t_\textrm{bs} - \Delta t_\textrm{bs}, t_\textrm{bs} + \Delta t_\textrm{bs}$] (user-defined), 
	\item Calculate a robust estimate of the average magnetic field, e.g. the median of the magnetic field in half of this interval, noted $|B_\textrm{sw}|_{1/2}$. By definition $|B_\textrm{sw}|_{1/2}$ corresponds to the assumed solar wind region magnetic field. For crossings from the solar wind into the sheath, the first half-interval is selected. For crossings from the sheath to the solar wind, the second half-interval is selected. There are two possibilities at this junction: 
	\begin{enumerate}
	    \item If $|B_\textrm{sw}|_{1/2} > B_\textrm{th}$, where $B_\textrm{th}$ is the solar wind-to-magnetosheath threshold (user-defined), the position of the solar wind region is difficult to assess. In that case:
	    \begin{enumerate}
		    \item Make the time interval float around the estimated location of the shock, by increments of $\Delta t_\textrm{bs}/3$ in one direction or the other, so that $t_\textrm{bs} = t_\textrm{bs} \pm \Delta t_\textrm{bs}/3$,
		    \item Repeat interval shift until $|B_\textrm{sw}|_{1/2} \leq B_\textrm{th}$ or until a maximum shift of $2\Delta t_\textrm{bs}$ is effected from the estimated shock timing in either direction. If $|B_\textrm{sw}|_{1/2} > B_\textrm{th}$ still, either the spacecraft is always in the sheath or in an usually high solar wind $B$-field region, in which case the crossing is altogether ignored and removed from the database. If not, go to step (b):
	    \end{enumerate}	
	    \item If $|B_\textrm{sw}|_{1/2} \leq B_\textrm{th}$, this half-interval is a good candidate for undisturbed solar wind conditions. In that case:
	    \begin{enumerate}
		    \item Calculate the running Median Absolute Deviation (MAD) $\varsigma_\textrm{mad,B}$ of the total $B$-field signal over a temporal window of duration $T$ (user-defined) in the chosen half-interval, and smooth further the result with a running median over a time span enclosing the shock structure in its entirety (for example $2T$ or $3T$). This also helps remove potentially abrupt but temporally isolated changes in the signal. Note that this particular choice of $\varsigma_\textrm{mad,B}$ is somewhat arbitrary. After several tests including running standard deviations, normalised or not to the "solar wind" signal, the choice of a smoothed running MAD was empirically found to work consistently well with the MAVEN dataset at \qty{1}{s} resolution. For all times $t$ at which the total magnetic field $B_\textrm{tot}$ is measured over a running interval $[t_i, t_i+T]$:
		    \begin{linenomath}
		    	\begin{align}
				\varsigma_{\textrm{mad,B}}(t) = \left\langle \mathrm{median}\left(\left|B_\textrm{tot}(t_i) - \underset{T}{\mathrm{median}}(B_\textrm{tot})\right|\right)\right\rangle \quad \forall t \in \left[t_i,t_i+T\right]
			\end{align}
		    \end{linenomath}
		    \item Compare $\varsigma_\textrm{mad,B}$ to threshold value $\varsigma_{\textrm{th}}$ (user-defined). A jump above a certain threshold $\varsigma_{\textrm{th}}$ indicates a transition between a less turbulent region ($\varsigma_\textrm{mad,B} < \varsigma_{\textrm{th}}$) to a more turbulent one ($\varsigma_\textrm{mad,B} > \varsigma_{\textrm{th}}$), an indicator of the presence of a shock-like structure. If this threshold is reached, take the first (respectively, last) time this happens in the chosen interval for solar wind-to-sheath (respectively, sheath-to-solar wind) crossings, and correct the original timing $t_\text{bs}$ of step $1$ to new $t_\textrm{bs,corr}$. If not, discard crossing.
	    \end{enumerate}
	    \item Repeat for each orbit.
	\end{enumerate}
\end{enumerate}

At Mars, we tested step $1$ (predictor) in the previous section with the MAVEN dataset: on average, the detected shock was within \qty{\pm0.08}{$R_p$} (\qty{\pm270}{km}) of the true shock crossing, corresponding to about \qty{\pm30}{min} of data along the orbit. We thus set the interval of study for the corrector algorithm to \qty{\Delta t_\textrm{bs}= 30}{min}. This value depends on the orbit inclination with respect to the bow shock surface and is thus mission-dependent.

We determined the typical user-defined threshold values for the MAVEN mission manually, for simplicity, on a reduced dataset. We found a good compromise by trial and error with $B_\textrm{th}$\,$=$\,$11$\,nT, because the undisturbed solar wind magnetic field in Mars' vicinity is of the order of $2$--$6$\,nT on average \cite{slavin_solar_1981}, but can reach up to about \qty{10}{nT} or more when solar transient effects such as coronal mass ejections or co-rotating interaction regions are involved \cite{liu_statistical_2021}. In future studies, a more dynamic criterion in step (3a) may be preferred, i.e.  where the amplitude of the magnetic field is normalised to the assumed upstream solar wind value. The criterion for being in the magnetosheath could for example become $\gamma$\,$=$\,$|\mathbf{B}|/ |B_\textrm{sw}|_{1/2}$\,$>$\,$\gamma_\textrm{th}$, where $\gamma_\textrm{th}$ is an adequately chosen threshold ($\gamma$\,$\sim$\,$1.5$ for a clear increase of magnetic field when moving into the magnetosheath, with nominal solar wind levels $\gamma$\,$\sim$\,$1$). 

Because the shock appears in measurements as a turbulent structure whereas the solar wind is on average less so, step (3b-i) calculates a measure of the variability of the magnetic field in the vicinity of the shock. The value of $T$ is also mission- and instrument-dependent; for MAVEN/MAG data at a resolution of \qty{1}{s}, a value \qty{T=120}{s} was chosen, which adequately captures the mean magnetic field variations. When in the solar wind, we found that the intrinsic variability of the signal $\varsigma_\textrm{mad,B}$, calculated over a running window of duration $T=120$\,s, was on average less than $0.5$, which is adopted as the threshold $\varsigma_{\textrm{th}}$. This makes it possible to detect the very first perturbations in the solar wind leading to the creation of the shock structure, a point which is identified here as the position of the shock proper at time $t_\textrm{bs}$. Again, in future studies, the threshold can be normalised to the magnetic field level, as in \citeA{halekas_structure_2017} where, together with constraints on plasma parameters, the normalised root-sum-squared value of the magnetic field was chosen so that $\text{RSS}(\mathbf{B})/|\mathbf{B}| < 0.15$ to identify undisturbed solar wind intervals.

{\bf Application to MAVEN dataset.}
Applied to the MAVEN dataset at 1-s orbital and magnetic field resolution, the corrector algorithm reaches an accuracy of \qty{\pm0.02}{$R_p$} (\qty{\pm70}{km}) around the ``true" shock (manually picked on a reduced dataset for comparison), a factor $4$ increase in accuracy with respect to step $1$. In the temporal datasets, this corresponds to only a few minutes of continuous data along MAVEN's orbit. This is epitomised in Figure\ \ref{fig:BowShockExamples}, which shows examples of shock crossing predictions from step $1$ (dashed lines) in magnetic field data (left axis) as compared to  the predictions from the predictor-corrector algorithm (unbroken lines, red for crossings into the magnetosheath, and blue for crossings into the solar wind). Right axes in green show $\varsigma_\textrm{mad,B}$ and the associated threshold of $0.5$, where $\varsigma_\textrm{mad,B}<0.5$ mainly occurs for upstream solar wind intervals. With the addition of criterion $B_\textrm{th}$ on the total magnetic field, only clear solar wind regions are captured by our algorithm, whereas ambiguous regions from the point of view of the magnetic field data are rejected.

At the beginning of the mission (Figure\ \ref{fig:BowShockExamples}a, Dec. 2014), the automatic predictor algorithm gives a reliable estimate of this $q_\perp$ shock's position: this is expected since the prediction is based on the 3D quadric model of \citeA{gruesbeck_three-dimensional_2018} who specifically performed shock fits on the first years of the mission. In this case, the predictor-corrector algorithm only corrects the shock's estimated location by a few minutes.
Figure\ \ref{fig:BowShockExamples}b (Sept. 2016) displays a case where the shock position is hard to ascertain from magnetic-field data only: the predictor estimate is off by up to $20$ minutes for all crossings. Because of the constraints on the magnetic field amplitude and the lack of significant variations in $|B|$ between 22:00 and 01:00\,UT, the predictor-corrector algorithm ignores the two first expected crossings but corrects well for the two next crossings (around 03:30 and 04:45\,UT).
Figure\ \ref{fig:BowShockExamples}c shows a more complex mix between sheath and solar wind conditions, and even though the boundaries are more subtle and the overall $B$-field magnitude below $10$\,nT, the predictor-corrector algorithm manages to estimate well the position of the $q_\parallel$ shock, ignoring possibly unclear crossings which do not fulfil the combined threshold conditions on $\varsigma_\textrm{mad,B}$ and $B_\textrm{th}$ (around 01:00\,UT).
Panel (d) shows yet another example of the superiority of the predictor-corrector algorithm for some very clear bow shock crossings in 2020, after the orbit of MAVEN had been altered into a different orbit than at the beginning of the mission.

The final corrected timings for the detections yield with this algorithm a lower estimate of the total actual number of crossings encountered by a spacecraft throughout its mission. Events occurring when $|B_\textrm{sw}|_{1/2} > B_\textrm{th}$, even after shifting the temporal window significantly, or when $\varsigma_\textrm{mad,B}<\varsigma_{\textrm{th}}$, were discarded in the final selection as can be seen in Figure\ \ref{fig:BowShockExamples}. They may indicate that the magnetic field was either too turbulent or too complex in its structure (e.g. multiple crossings as is regularly the case with $q_\parallel$ crossings) for the corrector algorithm to capture. Moreover, as discussed previously, the analytical approximation model used for the determination in step $1$ (either 2D or 3D) is likely to underestimate the true number of crossings due to the planetary bow shock variability \cite{halekas_structure_2017}. Because our study is primarily interested in the statistical position of the bow shock throughout the mission, this loss of potential detections may be compensated by the large number of orbits of the considered spacecraft.

For MAVEN, from the original $16,515$ candidate detections from step 1 (predictor) using the 3D model of \citeA{gruesbeck_three-dimensional_2018} as a first approximation, our predictor-corrector algorithm selected $14,929$ events ($7,494$ detections from the solar wind to the magnetosheath and $7,435$ from the sheath to the solar wind) for the period 01 November 2014 to 07 February 2021. This is a $10\%$ decrease in number of crossings, leaving out the less ambiguous events from the predictor algorithm. On average, the correction to the original timing is within about \qty{\Delta t\pm 20}{min} along the spacecraft orbit, which corresponds to a percentage difference in radial distance of about $\pm 25\%$ (not shown). 
The calculated average difference $|\Delta R_{bs}|$ over the entire dataset between predictor and predictor-corrector algorithm is \qty{\sim 0.11}{$R_p$}, i.e. \qty{\sim 380}{km} or about $5\%$ difference. The complete list of crossings (times and spatial coordinates) is compiled in \citeA{simon_wedlund_2021}.

\begin{figure}[htb]
 \includegraphics[width=0.75\columnwidth]{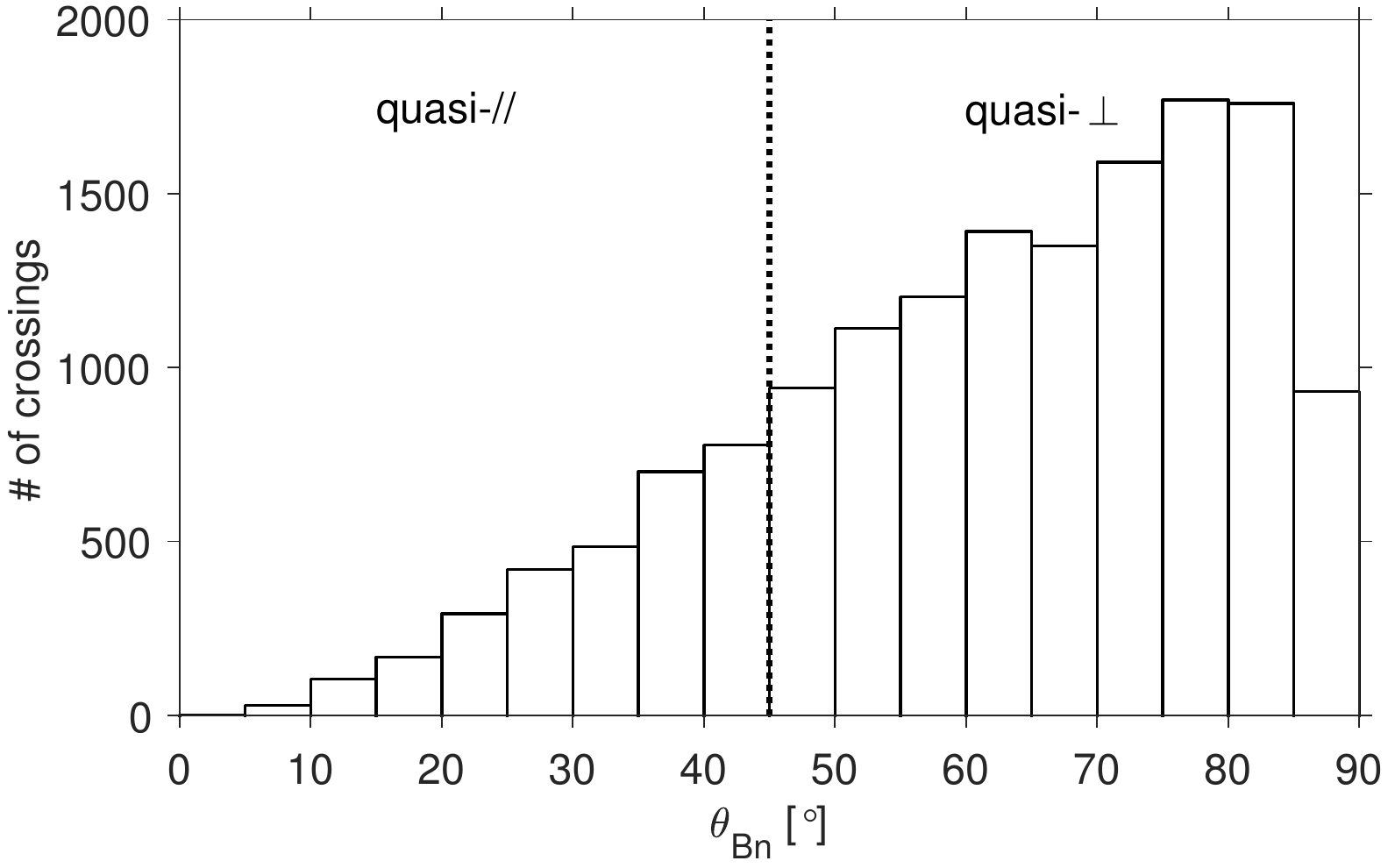}
 \caption{Statistical distribution of crossings with respect to $\theta_{B{\rm n}}$ angles, the angle between the normal to the shock and the average magnetic field direction.  The limit between $q_\parallel$ and $q_\perp$ conditions is for $\theta_{B{\rm n}} = 45\deg$. \label{fig:thetaBn}
 }
\end{figure}

Out of these $14,929$ crossings, we found that an overwhelming number ($11,967$) was predominantly $q_\perp$ crossings with $\theta_{B{\rm n}} > 45\deg$ as calculated with formula\,(\ref{eq:normalCartesian3D}), the remaining $2,962$ events being classified as more reminiscent of $q_\parallel$ shock conditions. It is important to recall that the method is biased towards clear signatures of $q_\perp$-like bow shock crossings, in practice filtering out many $q_\parallel$ crossings, which may explain part of the discrepancy.  
About $22\%$ of the $q_\perp$ crossings are highly perpendicular shocks ($\theta_{B{\rm n}} \geq 80\deg$, $2,674$ crossings), whereas only $\sim1\%$ of the $q_\parallel$ shocks are highly parallel ones ($\theta_{B{\rm n}} \leq 10\deg$, $30$ crossings).  This is shown in Figure\ \ref{fig:thetaBn}, where the highest number of crossings occurs for $\theta_{B{\rm n}}\sim 80\deg$. This is also in qualitative agreement with the results of \citeA{vignes_factors_2002} for the MGS mission, when they investigated a proportion of $93$ $q_\perp$ shocks for only $23$ $q_\parallel$ shocks.

\section{Application to the study of the Martian bow shock variability}

In this section, as an application and statistical test of our new predictor-corrector algorithm, we present 2D and 3D fits of the average bow shock position with the MAVEN spacecraft for the 01 November 2014--07 February 2021 period. First, we sort out the detected shock positions \cite<compiled in>{simon_wedlund_2021} by Mars year (MY32 to MY35 included), aerocentric solar longitude Ls range (four seasons centred on equinoxes and solstices), EUV flux (two regimes, one for higher solar flux and lower solar flux), and shock geometry ($q_\perp$ or $q_\parallel$). These cases correspond to:
\begin{itemize}
	\item Mars years 32 (incomplete), 33, 34 and 35, inspired by the work of \citeA{hall_martian_2019} on MEX datasets,
	\item Solar longitude Ls ranges from $[315\deg$--$45\deg]$ (centred on Northern Hemisphere [NH] spring equinox), $[45\deg$--$135\deg]$ (NH summer solstice), $[135\deg$--$225\deg]$ (NH autumn equinox), $[225\deg$--$315\deg]$ (NH winter solstice). Ls defines the geographic Martian season, with $\text{Ls} = 251\deg$ ($\text{Ls} = 71\deg$, respectively) marking perihelion (aphelion) conditions. 
	\item Two EUV flux levels, inspired by the works of \citeA{halekas_structure_2017} and \citeA{gruesbeck_three-dimensional_2018} on the early MAVEN datasets. The EUV flux at Mars is obtained using the FISM-IUVS daily irradiances at \qty{121.5}{nm} calculated from the Mars EUVM model \cite{thiemann_maven_2017}. The median of the EUV flux in the 2014-2021 period is \qty{0.0028}{W/m$^2$} and defines two EUV flux levels, one "high" for fluxes above that limit, one "low" for fluxes below.
\end{itemize}
In Section\ \ref{sec:statisticalPosition}, we perform 2D and 3D fits to the found bow shock positions using the spacecraft ephemerides. A discussion of these fits and what they imply is given in Section\ \ref{sec:discussion}.
 
\subsection{Statistical position of the Martian bow shock}\label{sec:statisticalPosition}
{\bf 2D case}. In 2D MSO aberrated coordinates, the polar equation, Equation\,(\ref{eq:polarBSmodel}), can be rewritten in the linear form $y = ax + b$ \cite<see>[]{trotignon_martian_2006}:
\begin{linenomath}
\begin{align}
 	r = L - \epsilon \left(X^\prime-x_F\right) \label{eq:fit2Dconic},
\end{align}
\end{linenomath}
with a linear regression in the $(r, X^\prime-x_F)$ space performed for a chosen focus location $x_F$. First we chose a focus location randomly between 0 and \qty{1}{$R_\textrm{p}$}, and for each linear fit performed, the residuals are calculated. The adopted focus point is the one that minimises the residuals. Because MAVEN's orbits are not suited to bow shock detections for $X^\prime < -0.5\,R_\textrm{p}$, additional constraints on the tail distributions are necessary to obtain a more realistic conic fit. This can be achieved for example by using the predictions from a chosen pre-existing model for deeply negative $X^\prime$ values, such as those of \citeA{edberg_statistical_2008} (noted 'E08' in the following) or \citeA{hall_martian_2019} (noted 'H19') where bow shock detections were reported downstream to $X^\prime_\textrm{min} \approx -1.5\,R_\textrm{p}$ and to $\approx -5\,R_\textrm{p}$, respectively. First, additional "ghost" points (representing $10\%$ of the total number of detections for the considered case) are calculated for $X^\prime_\textrm{min} < X^\prime < -0.5\,R_\textrm{p}$ for the chosen model and randomised spatially around this result to give a more realistic tail spread. The linear regressions are then performed on the new constrained dataset. Tests were performed on the robustness of this method using different analytical models: E08 and H19 fits are essentially the same around the nose of the shock downstream to about $-1\,R_\textrm{p}$ where patent differences start to appear. Because of the added cloud of ghost points, this is expected and thus 2D fits presented below are only valid in practice in the range $[-0.5\,R_\textrm{p},\ R_\textrm{ss}]$. Incidentally, differences on the terminator and subsolar standoff distances are less than $<2\%$ in each case. It is noteworthy to add that the determination of the nature of the conic section found from the fits can be significantly altered when using tail models either from E08 (hyperbola) or from H19 (fits' nature given by these authors, ranging from ellipse to hyperbola, depend on the Mars year considered): in that case, the fit's nature will naturally be biased towards matching that of their respective parent tail model.

\begin{figure*}[htb]
 \includegraphics[width=\columnwidth]{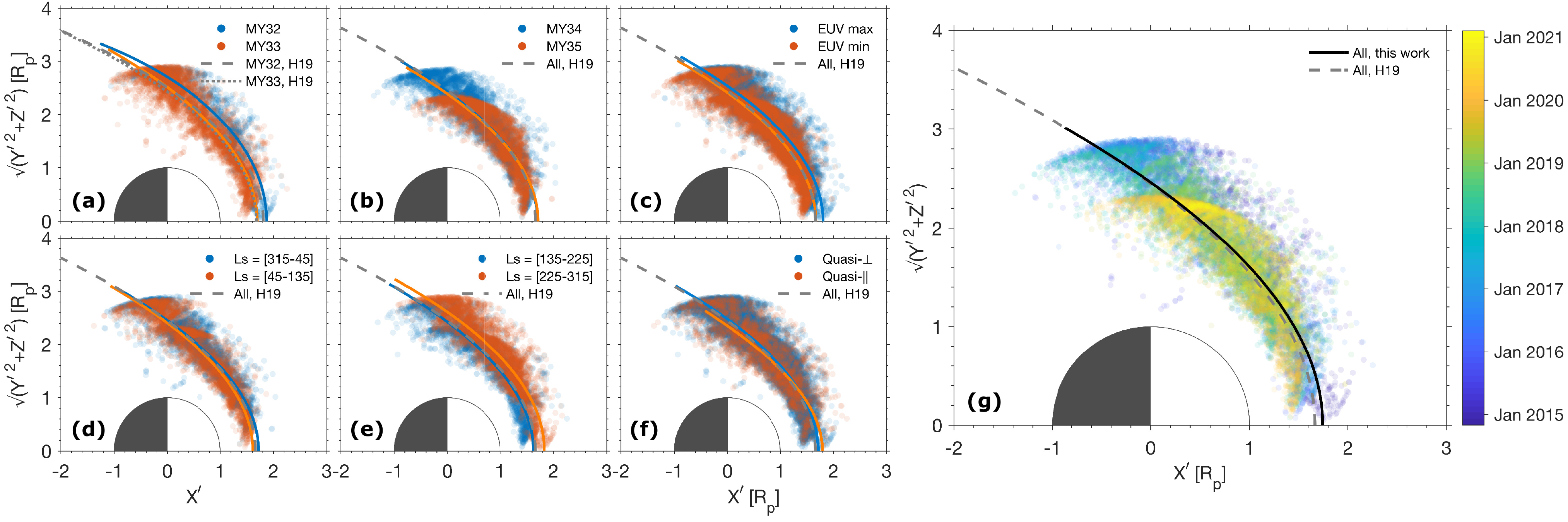}
 \caption{2D fits performed on the predictor-corrector algorithm for the detection of bow shock crossings in the MAVEN dataset, $2014-2021$ in aberrated MSO coordinates ($X^\prime_\textrm{MSO},Y^\prime_\textrm{MSO}, Z^\prime_\textrm{MSO}$), and parametrised in Table\ \ref{tab:Conic2Dparameters}. Panels (a) and (b): vs Mars Years $32$ to $35$. Panels (c) and (d): vs Ls (season) ranges. Panel (e): vs EUV flux levels. Panel (f): vs shock geometry ($q_\perp$ and $q_\parallel$). Panel (g): all detected points in the current database colour-coded by year, and comparison to the analytical quadric fit of \citeA{hall_martian_2019}. All coordinates are expressed in units of the planet's radius, i.e. $R_\textrm{p} = 3389.5$\,km. Superimposed on all panels are the corresponding analytical models of \citeA{hall_martian_2019} for Mars years $27$--$33$, except for Mars years 32 and 33, where their corresponding yearly fits are plotted. Candidate detections points for each case are also drawn as filled circles of varying colours, with the opacity giving a measure of the density of points in that area, giving more or less weight to the fitting method. \label{fig:2DFits}
 }
\end{figure*}

Table\ \ref{tab:Conic2Dparameters} and Figure\ \ref{fig:2DFits} display our 2D fits when the E08 tail model supplements the MAVEN dataset for additional constraints on the tail. Candidate bow shock detections are also drawn as semi-transparent circles. Despite the constraints on the predictor-corrector algorithm, several detected points appear to fall well into the magnetosheath of Mars, and are false detections. Because of their relative scarcity and thanks to the large statistical database, these points do not significantly impact the final fits, which remain robust.

\begin{table*}[t]
    \centering
    \caption{Martian bow shock 2D conic parameters in aberrated MSO coordinates from linear regression fits applied to Equation\,(\ref{eq:fit2Dconic}) and the MAVEN orbits and magnetic field data (predictor-corrector algorithm). Subsolar and terminator standoff distances $R_\textrm{ss}$ and $R_\textrm{td}$ are calculated with Equations\,(\ref{eq:standoffDistanceConic}) and (\ref{eq:standoffTerminatorDistanceConic}). For hyperbolae, the Mach cone aperture $\varrho$ is also given as calculated by Equation\,(\ref{eq:MachConeAngle}). For each fit, the coefficient of determination $R^2$ gives a measure of the goodness of the linear regression. Due to the large data spread, uncertainties on $R_\textrm{ss}$ and $R_\textrm{td}$ are of the order of $5\%$ and of the order of $2\%$ for the other quantities.
    }
    \footnotesize
    \begin{tabular}{l | c c c c c c | l c | r}
        {\bf Case} & $\varepsilon$ & $L [R_\textrm{p}]$ & $x_F [R_\textrm{p}]$ & $R_\textrm{ss} [R_\textrm{p}]$ & $R_\textrm{td} [R_\textrm{p}]$  &  $R^2$ & Nature & $\varrho [\deg]$ & $\#$ detections\\
        \hline
        \emph{All points, this work} & 1.00 & 1.75 & 0.86 & 1.74 & 2.46 & 0.98 & Parabola & $-$ & 14929\\
     MY32, this work & 0.83 & 2.24 & 0.65 & 1.87 & 2.70 & 0.97 & Ellipse & $-$ & 1196\\
     MY33, this work & 0.99 & 1.88 & 0.75 & 1.69 & 2.51 & 0.98 & Ellipse & $-$ & 4586\\
     MY34, this work & 1.02 & 1.72 & 0.84 & 1.69 & 2.44 & 0.96 & Hyperbola & 11 & 5073\\
     MY35, this work & 1.02 & 1.63 & 0.91 & 1.72 & 2.39 & 0.98 & Hyperbola & 11 & 4074\\
     Ls = [$315^\circ-45^\circ$], this work & 1.01 & 1.73 & 0.86 & 1.72 & 2.45 & 0.98 & Hyperbola & 8 & 3793\\
     Ls = [$45^\circ-135^\circ$], this work & 1.00 & 1.81 & 0.71 & 1.61 & 2.42 & 0.99 & Parabola & $-$ & 3746\\
     Ls = [$135^\circ-225^\circ$], this work & 0.99 & 1.82 & 0.71 & 1.62 & 2.42 & 0.98 & Ellipse & $-$ & 3134\\
     Ls = [$225^\circ-315^\circ$], this work & 0.98 & 1.91 & 0.86 & 1.82 & 2.62 & 0.98 & Ellipse & $-$ & 4256\\
     EUV flux $\geq 0.0028$\,W/m$^2$ & 1.00 & 1.79 & 0.91 & 1.80 & 2.54 & 0.98 & Parabola & $-$ & 6502\\
     EUV flux $< 0.0028$\,W/m$^2$ & 1.00 & 1.75 & 0.79 & 1.67 & 2.41 & 0.98 & Parabola & $-$ & 8427\\
     Quasi-$\perp$ & 1.00 & 1.79 & 0.82 & 1.72 & 2.48 & 0.98 & Parabola & $-$ & 11967\\
     Quasi-$\parallel$ & 1.06 & 1.47 & 1.07 & 1.78 & 2.37 & 0.94 & Hyperbola & 19 & 2962\\
        \hline   
    \end{tabular}
    \label{tab:Conic2Dparameters}
\end{table*}

{\bf 3D case}. In 3D, we perform quadric fits using the method first put forward by \citeA{taubin_estimation_1991}, adapted to the quadratic surface of Equation\,(\ref{eq:quadraticBS}). This fitting method constructs scatter matrices from local gradients $\mathbf{S}$ of tested model $\mathbf{T}$ and finds the diagonal matrix of the generalised eigenvalue problem so that $\mathbf{T}\mathbf{v} = \lambda \mathbf{S} \mathbf{v}$, where $\mathbf{v}$ is the generalised eigenvector of $\mathbf{T}$ and $\mathbf{S}$, and $\lambda$ are the eigenvalues. Because of the scatter of points in the database, uncertainties on the found parameters $A$ to $I$ are of the order of $1\%$, in a least-squares sense.

\begin{figure*}[tb]
 \includegraphics[width=0.7\textwidth]{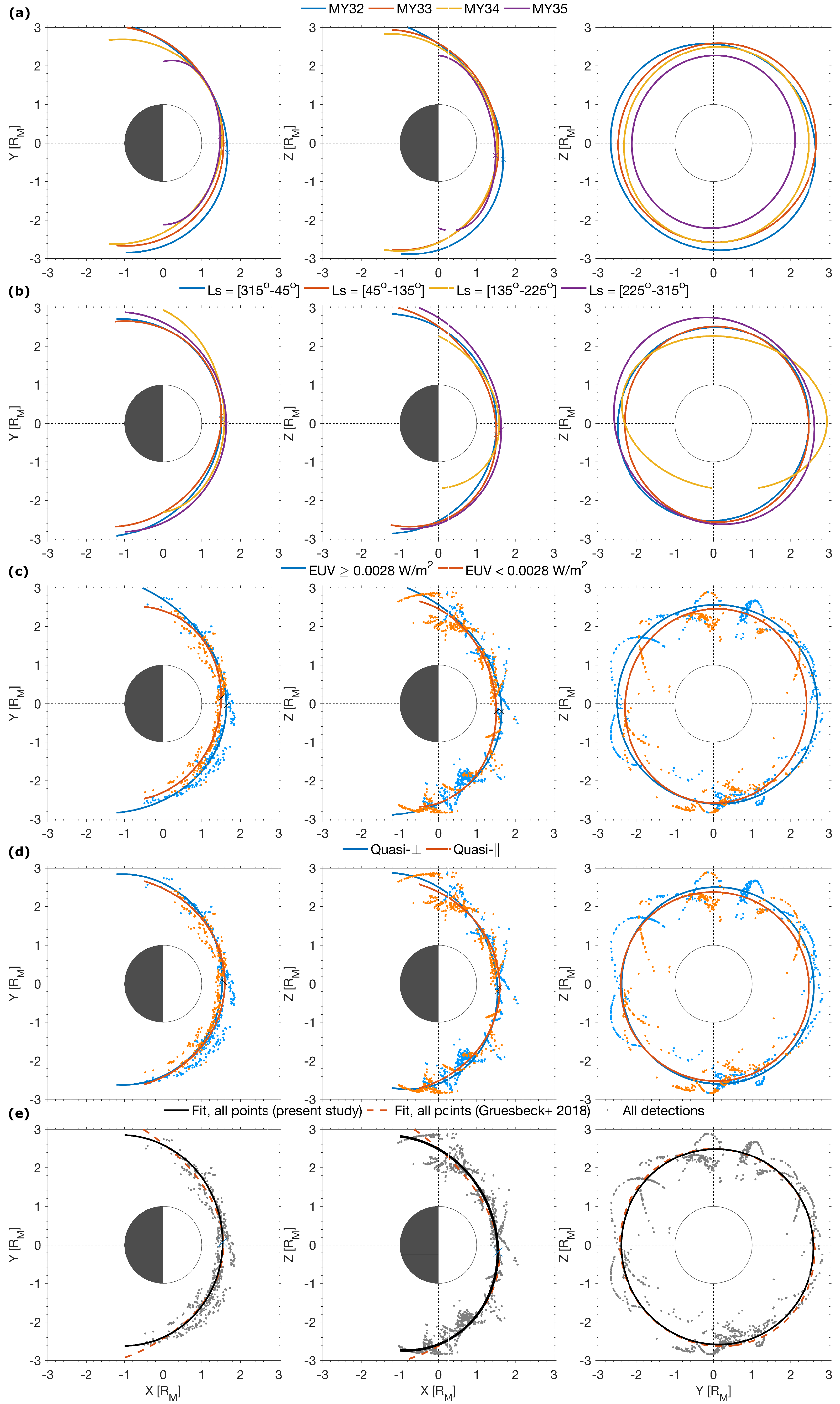}
 \caption{3D fits performed on the refined predictor-corrector algorithm for the detection of bow shock crossings in the MAVEN dataset, $2014-2021$ in the $X_\textrm{MSO}-Y_\textrm{MSO}$, $X_\textrm{MSO}-Y_\textrm{MSO}$ and $Y_\textrm{MSO}-Z_\textrm{MSO}$ planes (traces of ellipsoids of revolution parametrised in Table\ \ref{tab:Conic3Dparameters}). (a) vs Mars Years. (b) vs Ls (season) ranges. (c) vs EUV flux levels, with their corresponding subset of detected points (blue and orange dots). (d) vs bow shock geometry, $q_\perp$ (blue dots) and $q_\parallel$ (orange dots). (e) all detected points in the current database with a comparison of present fit (black line) to the analytical quadric fit of \citeA{gruesbeck_three-dimensional_2018} (orange dashed line). On each figure, superimposed crosses show where the nose of the shock is located, in the plane of projection (see \ref{sec:appendix2}). All coordinates are expressed in units of the planet's radius, i.e. $R_\textrm{p} = 3389.5$\,km.
 \label{fig:3DFits}
 }
\end{figure*}

Table\ \ref{tab:Conic3Dparameters} collects all 3D fit parameters for each case; all fitted surfaces are ellipsoids of revolution. For completeness, we present and give the physical interpretation of these parameters in \ref{sec:appendix1}, in terms of principal axes, their direction and lengths and the centering of the ellipsoids.
Figure\ \ref{fig:3DFits} shows the corresponding fits and their sections onto the $X_\textrm{MSO}-Y_\textrm{MSO}$ (dawn-dusk hemispheres), $X_\textrm{MSO}-Z_\textrm{MSO}$ (South-North hemispheres) and $Y_\textrm{MSO}-Z_\textrm{MSO}$ (at the terminator, i.e. $X_\textrm{MSO}=0$) non-aberrated MSO coordinates. 

Because of the spacecraft changing orbits during the mission, some of the ellipsoid fits appear anomalous in their orientation. This is especially obvious for MY35 when MAVEN, as of Mars 2020, decreased its apogee to $\sim4500$\,km and hence its revolution period to $3.5$\,h to accommodate Mars 2020 rover operations on the ground. Consequently, MAVEN only seldom explored regions below $X < 0\,R_p$ for half of MY35: this makes it difficult to constrain the fit, and we end up with an ellipsoid having its longest principal axis unphysically tilted almost $90\deg$ in the $X$-$Z$ plane (panel a, purple curves of Figure\ \ref{fig:3DFits}). A similar issue is found for $\text{Ls}=135$--$225\deg$ (panel b, yellow, Figure\ \ref{fig:3DFits}), which has the lowest number of detections among Ls ranges and for which the orbit was never favourable for detections due to orbit precession. Thus, in these two cases, no physical interpretation should be drawn from the axes orientations of the ellipsoid and the fit should only be valid for near-subsolar crossings. More robust physics-based analytical models could be used to overcome these fitting issues \cite{kotova_physics-based_2021}.

\begin{table*}[t]
    \centering
    \caption{Martian bow shock 3D conic parameters from quadric surface fits applied to the MAVEN orbits and magnetic field data (predictor-corrector algorithm). See Equation\,(\ref{eq:quadraticBS}) for the definition of parameters $A$ to $I$ and Equations\,(\ref{eq:Rss}), (\ref{eq:Rtdy}) and (\ref{eq:Rtdz}) for those of the subsolar standoff distance along the $X_{\rm MSO}$ axis and the terminator standoff distances along the $Y_{\rm MSO}$ and $Z_{\rm MSO}$ axes. Uncertainties on the parameters are of the order of $1\%$ in a least squares sense. All quadrics below are ellipsoids. The domain of validity for each fit is shown in Figure\ \ref{fig:3DFits}: fits are valid for $X_{\rm MSO}\geq-0.5\,R_p$ on average. The number of fitting points used for each case is the same as for the 2D fits, see Table\ \ref{tab:Conic2Dparameters} (last column). Also, see \ref{sec:appendix1} for a physical interpretation of the tabulated parameters.
    }
    \scriptsize
    \begin{tabular}{l c c c c c c c c c c c c}
        {\bf Case} & $A$ & $B$ & $C$ & $D$ & $E$ & $F$ & $G$ & $H$ & $I$ & $R_\textrm{ss}$ & $R_{\textrm{td},y}$ & $R_{\textrm{td},z}$ \\
        \hline
        \citeA{gruesbeck_three-dimensional_2018}$^{a}$ & 0.0490 & 0.1570 & 0.1530 & 0.0260 & 0.0120 & 0.0510 & 0.5660 & -0.0310 & 0.0190 & 1.557 & 2.624 & 2.495 \\
     \emph{All points}, this work & 0.1769 & 0.1609 & 0.1559 & 0.0057 & 0.0044 & 0.0281 & 0.3773 & -0.0323 & 0.0143 & 1.539 & 2.595 & 2.487 \\
     MY32, this work & 0.1369 & 0.1419 & 0.1400 & 0.0381 & 0.0178 & 0.0547 & 0.3783 & 0.0044 & 0.0279 & 1.654 & 2.639 & 2.575 \\
     MY33, this work & 0.1660 & 0.1516 & 0.1501 & 0.0235 & -0.0061 & 0.0203 & 0.3807 & -0.0284 & -0.0018 & 1.562 & 2.664 & 2.587 \\
     MY34, this work & 0.1719 & 0.1742 & 0.1551 & -0.0112 & -0.0068 & 0.0165 & 0.3955 & -0.0256 & 0.0127 & 1.522 & 2.470 & 2.499 \\
     MY35, this work & 0.5577 & 0.2245 & 0.2000 & -0.0509 & -0.0103 & 0.0963 & -0.1421 & 0.0011 & -0.0119 & 1.472 & 2.108 & 2.266 \\
     Ls = [$315^\circ-45^\circ$], this work & 0.1554 & 0.1625 & 0.1587 & -0.0265 & -0.0081 & 0.0023 & 0.4260 & 0.0013 & 0.0044 & 1.513 & 2.477 & 2.496 \\
     Ls = [$45^\circ-135^\circ$], this work & 0.1719 & 0.1761 & 0.1555 & -0.0275 & 0.0048 & 0.0564 & 0.4107 & -0.0287 & 0.0057 & 1.497 & 2.466 & 2.518 \\
     Ls = [$135^\circ-225^\circ$], this work & 0.2490 & 0.1473 & 0.2624 & 0.0409 & 0.0415 & 0.1098 & 0.2368 & -0.0941 & -0.1510 & 1.584 & 2.945 & 2.261 \\
     Ls = [$225^\circ-315^\circ$], this work & 0.1559 & 0.1484 & 0.1400 & 0.0047 & 0.0227 & 0.0400 & 0.3583 & -0.0072 & -0.0197 & 1.632 & 2.620 & 2.744 \\
     EUV flux $\geq$ 0.0028\,W/m$^2$ & 0.1096 & 0.1480 & 0.1500 & 0.0274 & 0.0031 & 0.0355 & 0.4329 & -0.0293 & 0.0045 & 1.634 & 2.700 & 2.567 \\
     EUV flux $< 0.0028$\,W/m$^2$ & 0.2138 & 0.1807 & 0.1577 & -0.0231 & -0.0051 & 0.0314 & 0.3473 & -0.0207 & 0.0185 & 1.498 & 2.410 & 2.460 \\
     Quasi-$\perp$ & 0.1798 & 0.1607 & 0.1539 & 0.0016 & 0.0040 & 0.0330 & 0.3777 & -0.0348 & 0.0124 & 1.531 & 2.605 & 2.509 \\
     Quasi-$\parallel$ & 0.1427 & 0.1675 & 0.1666 & -0.0004 & 0.0051 & 0.0050 & 0.3992 & -0.0098 & 0.0230 & 1.595 & 2.473 & 2.382 \\
        \hline 
        \multicolumn{13}{l}{$^{a}$For \emph{all points} considered in their data subset.}
    \end{tabular}
    \label{tab:Conic3Dparameters}
   
\end{table*}

\subsection{Discussion}\label{sec:discussion}
Our 2D and 3D fits give some insight on how the Martian bow shock is moving globally for different conditions of Mars Year, Ls and EUV flux and complements previous studies with the MAVEN mission \cite{halekas_structure_2017,gruesbeck_three-dimensional_2018,Nemec2020}. As the bow shock position is connected to the balance between thermal pressure from the plasma in the ionosphere and the dynamic pressure from the solar wind, any variation of these two quantities will have repercussions on the position of the shock. 

When assuming axisymmetry around the aberrated axis $X^\prime$ in the 2D polar rectangular coordinates case, Table\ \ref{tab:Conic2DparametersPast} and the average subsolar and terminator distances can be a first guide for our interpretation. Our new results with MAVEN (all points, Table\ \ref{tab:Conic2Dparameters}) agree rather well with past measurements (Table\ \ref{tab:Conic2DparametersPast}) considering the data spread and estimated uncertainties: $R_\textrm{ss} = 1.74\pm0.09$ compared to $1.61\pm0.08\,R_\textrm{p}$ and $R_\textrm{td} = 2.46\pm0.13$ compared to $2.56\pm0.20\,R_\textrm{p}$. More specifically, for: 
\begin{itemize}
	\item Mars Years: the subsolar standoff distance decreases by as much as $10\%$ between MY32 and MY33--MY34, from $1.87$ to $1.69\,R_\textrm{p}$, although some of this variation may be stemming from the relatively lower statistics for the first year ($1,196$ points for MY32 compared to $>4,000$ for all other years), due to the MAVEN mission starting towards the end of MY32. A similar tendency is seen for terminator standoff distances, with a $11\%$ decrease seen between MY32 and MY35.
	Following \citeA{hall_martian_2019}, these variations may be connected through solar EUV irradiance to the solar cycle itself, when descending from the maximum of solar cycle $24$ (encompassed by MY32) towards a minimum of activity (MY34) and the start of solar cycle $25$ (MY35). A variation in $R_\text{ss}$ of similar magnitude ($\sim 7\%$ between minimum and maximum of activity) was shown by \citeA{hall_martian_2019} using MEX data for the previous solar cycles ($23$--$24$).
	\item Seasonal variations: in contrast, the Ls ranges have a much more even statistics throughout, with more than $3,000$ detections per season. Arguably, this makes comparing results between seasons statistically more significant than for the previous case. Overall, for northern spring equinox (Ls = $[315\deg-45\deg]$) and winter solstice (Ls = $[225\deg-315\deg]$) conditions, the bow shock appears to expand in the subsolar direction by about $7-13\%$ from its summer and autumn position ($R_\textrm{ss} \geq 1.72\,R_\textrm{p}$ compared to $R_\textrm{ss} \approx 1.61\,R_\textrm{p}$). Simultaneously, the area encompassed by the bow shock conic is also increased during those two instances.
	One possible driving factor behind these changes may be in turn linked to changes in Mars' dayside upper atmosphere and extended exosphere, and how they expand and contract with seasons, increasing or decreasing the size of the obstacle to the solar wind flow \cite<>[and references therein]{hall_annual_2016}. A denser lower atmosphere around perihelion ($\text{Ls}\sim 251\deg$ where the EUV flux is highest on average) and during the dust storm season in the autumn \cite{trainer_seasonal_2019} may drive the ionosphere to expand significantly at constant EUV flux \cite{sanchezcano_solar_2016,dubinin_expansion_2019}, offering a more efficient obstacle to the solar wind. Similarly the expansion of Mars' extended exosphere (notably modulated by the solar wind flux) increases the efficiency of the solar wind charge-exchange process \cite<with a net conversion of fast solar wind ions to slow-moving heavy ions of planetary origin, effectively slowing down the solar wind, see>{edberg_plasma_2009, halekas_structure_2017}. Both aspects result in the standoff distance moving outwards. The opposite effect is expected when upper atmosphere densities are lower in the deep summer and in the beginning of the autumn and the bow shock surface shrinks. The fits and characteristics of the shock appear consistent with this picture.
	\item EUV flux variations: the effect of a relatively larger flux on the shock position is twofold, globally increasing the ionisation rates in the ionosphere and through photoionisation of the extended exosphere as well as heating up and expanding the neutral atmosphere-exosphere system \cite{forbes_solar_2008,edberg_plasma_2009,hall_annual_2016}. Photoionisation of exospheric neutrals creates newly born ions that are picked up by the solar wind convective electric field, resulting in mass-loading and slowing down of the solar wind flow \cite<>[with the presence of pickup ions in the foreshock region]{yamauchi_seasonal_2015}. Such combined effects have been shown to expand the bow shock in the solar wind direction \cite{mazelle_bow_2004}. 
	The two fits we present here, one for higher and one for lower EUV fluxes (more than $6500$ points each), display the expected behaviour, with a larger standoff distance by $7\%$ and a noticeably larger flaring of the fitted conic for the higher EUV fluxes (terminator distances increasing from $2.41$ to $2.54\,R_\textrm{p}$ , i.e. $5\%$).
	\item Shock conditions: $q_\parallel$ and $q_\perp$ bow shock crossings are related to the average interplanetary magnetic field's (IMF) direction and the spacecraft's orbit (more precisely, the spherical quadrant in which the spacecraft emerges into the solar wind). Because the predictor-corrector algorithm favours $q_\perp$ detections (Section\ \ref{sec:PredictorCorrector}), the statistics between the two cases is heavily unbalanced \cite<see>[for a similar result]{vignes_factors_2002}. On average, we find no significant difference between the two conditions, with the shock surface slightly contracting and flaring up in $q_\perp$ conditions with respect to $q_\parallel$ conditions ($|\Delta R_\text{ss,td}| \sim 4\%$). Such a tendency is marginal considering that these percentages are at the precision limit obtained with the fits.

\end{itemize}

Let us now look at our 3D fit results. Figure\ \ref{fig:3DFits} clearly shows several asymmetries depending on the Mars year, Ls, EUV flux and shock condition. The usual pronounced North-South asymmetry ($X_{\rm MSO}-Z_{\rm MSO}$ plane, second column, and also $Y_{\rm MSO}-Z_{\rm MSO}$ plane, third column), mostly ascribed to the presence of crustal magnetic fields in the southern hemisphere \cite{gruesbeck_three-dimensional_2018}, is clearly seen for all cases with the standoff subsolar distances being skewed towards that hemisphere. This is shown by crosses representing the tip of the projected ellipsoid (calculated by the formulae in \ref{sec:appendix2}) located all in the fourth quadrant in the $X_{\rm MSO}-Z_{\rm MSO}$ plane. A similar tendency is sometimes marginally observed in the $X_{\rm MSO}-Y_{\rm MSO}$ plane (first column), when the shock surface is skewed towards the dawn hemisphere ($-Y_{\rm MSO}$), with standoff subsolar distances on average larger than on the dusk hemisphere. This is true for MY32 and MY33 (panel a) and for larger EUV fluxes (panel c). For lower EUV fluxes (and, incidentally, all other cases), the opposite seems to be taking place with the position of the maximum standoff distance being in the dusk $+Y_{\rm MSO}$ hemisphere. It is difficult at this stage to tell if these latter (small) effects may stem mainly or not from the dawn-dusk asymmetry of the atmosphere and hence of the ionosphere \citeA{gupta_dawn_2019}. Likewise, as in the 2D case, it is important to note that many drivers of the shock position (EUV flux, atmospheric seasons, etc.) all act in combination at any given time: our fits do not discriminate precisely between these effects. Consequently, a finer characterisation of each driver separately is left for another study.

\begin{figure*}[htb]
 \includegraphics[width=\columnwidth]{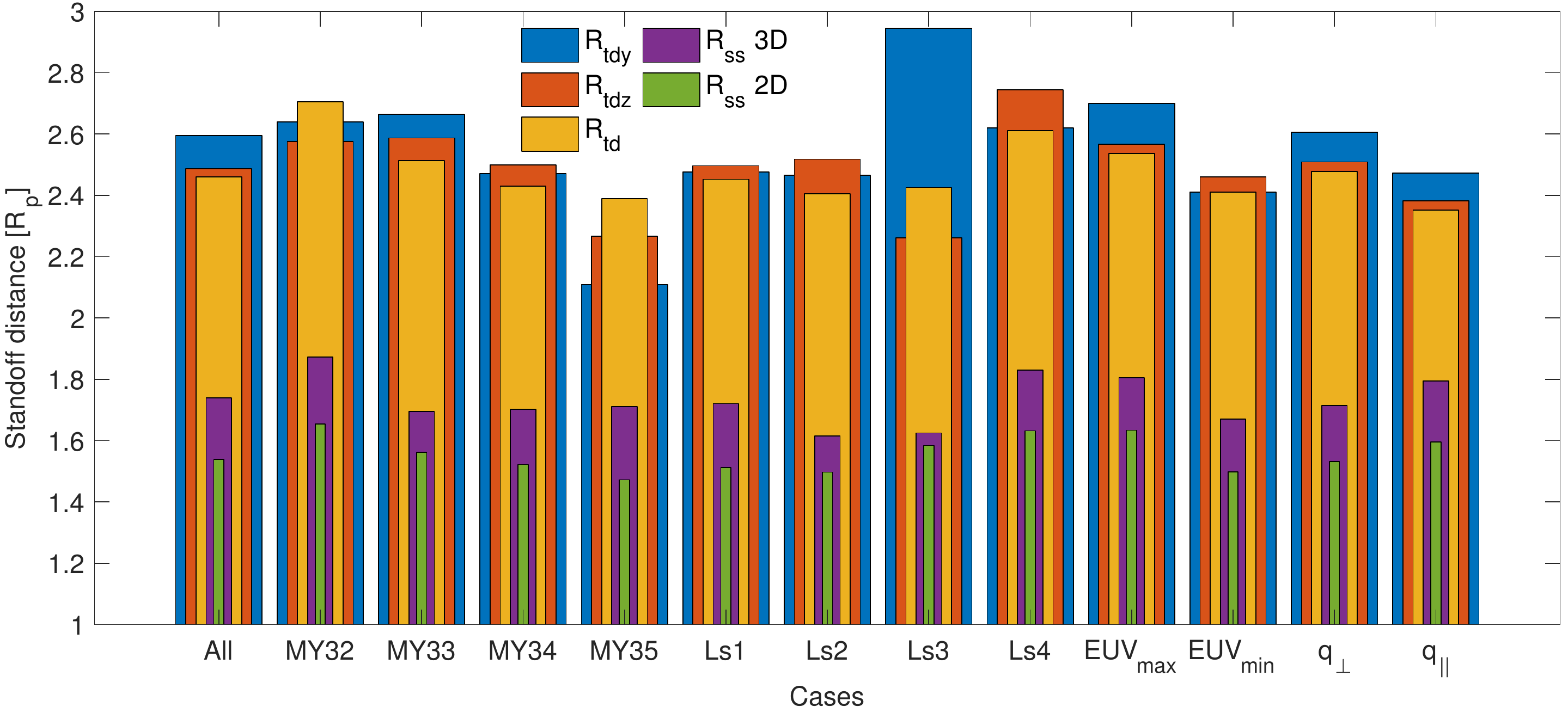}
 \caption{Comparison of standoff distances, both at the subsolar point and at the terminator, calculated from the 2D and 3D fits, and for each case as in Tables\,\ref{tab:Conic2Dparameters} and \ref{tab:Conic3Dparameters}. Terminator standoff values are in blue, orange and yellow (wider bars), whereas subsolar standoff values are in violet and green (thinner bars). For brevity in the axis labelling, Ls1 = $[315\deg$--$45\deg]$, Ls2 = $[45\deg$--$135\deg]$, Ls3 = $[135\deg$--$225\deg]$, Ls4 = $[225\deg$--$315\deg]$. All distances are expressed in units of the planet's radius, i.e. $R_\textrm{p} = 3389.5$\,km. \label{fig:compareStandoffs}
 }
\end{figure*}

A comparison between all standoff distances, subsolar and terminator alike, and calculated by our 2D and 3D algorithms is shown in Figure\ \ref{fig:compareStandoffs}, and based on Tables\,\ref{tab:Conic2Dparameters} and \ref{tab:Conic3Dparameters}. The standoff distances calculated from the 3D fits show the same tendencies as their 2D counterparts, although since the $X$ and $Y$ coordinates are not solar-wind aberrated and hence no axisymmetry is considered, the comparison between the 2D and 3D cases can only be that of general trends. From MY32 to MY35, a general decrease of standoff distances can be seen. Excluding $R_{{\rm td}y}$, the other standoff distances first steadily decrease from Ls centred on $0\deg$ (labelled ``Ls1'') to $180\deg$ (``Ls3'') but then increase significantly towards Ls values around $270\deg$ (``Ls4''), which may be linked to the EUV flux becoming maximum at perihelion $\text{Ls} = 251\deg$. This result is arguably in contrast to those presented in \citeA{vignes_factors_2002} although our statistics with MAVEN is much larger than in their study. In an identical way to the 2D fits, larger EUV fluxes result in a bow shock surface significantly expanding in the solar wind towards the subsolar direction. With respect to the geometry of the shock, the subsolar standoff distances $R_{\rm ss}$ appear to marginally increase from $q_\perp$ to $q_\parallel$ conditions, although the inverse trend is seen for the terminator distances. Again, these differences are slight, which may reflect in part the bias against $q_\parallel$ conditions of our bow shock estimator (thus yielding a low amount of $q_\parallel$ shock detections).

As a preliminary conclusion, we note that:
\begin{itemize}
	\item The $X_{\rm MSO}$--$Y_{\rm MSO}$ and $X_{\rm MSO}$--$Z_{\rm MSO}$ asymmetry seems particularly marked for Ls = $[135\deg$--$225\deg]$ (labelled ``Ls3'' on the figure), MY32, MY35 and higher EUV fluxes: it can readily be seen by comparing the length of the blue and red bars. As explained earlier, the number of points used for fits for MY32 is the lowest of all the cases because MAVEN arrived at Mars late in MY32. This is in qualitative agreement with the conclusions of \citeA{gruesbeck_three-dimensional_2018}. On average and outside of those special cases, the shock's shape stays rather symmetric about the $X_{\rm MSO}$ axis: the terminator distances $R_{{\rm td}z}$ (3D fits) and $R_{\rm td}$ (2D fits) indeed seem to match rather well most of the time. This axisymmetric tendency can be further amplified by aligning the $X_{\rm MSO}$--$Y_{\rm MSO}$ plane with the solar wind aberration system, rotating the 3D quadric surface $4$ degrees anticlockwise around the $Z_{\rm MSO}$ axis; new standoff distances for the 3D fits ($R_{{\rm td}z}$ and $R_{\rm ss}$) differ by less than $5\%$ with their corresponding 2D fits values (not shown). 
	\item Although the 3D and 2D conic fits retain strong similarities in their behaviour, the 3D fits (seemingly paradoxically) appear more robust and less affected by external assumptions. It is recalled here that not only do the 2D fits assume axisymmetry around $X^\prime_{\rm MSO}$, but certain 2D fits had to also be constrained at larger Euclidean distances from the centre of the planet due to the poor coverage of MAVEN for $X^\prime_{\rm MSO}<0.5\,R_\textrm{p}$. This superiority of the 3D fitting algorithm is due to: (i) the number of fitting variables ($A$ to $I$, allowing more flexibility despite risking over-determination of the linear system of equations), (ii) the natural asymmetry of the shock (albeit small), and (iii) the fitting points being statistically better distributed over a larger space (both in $X_{\rm MSO}$--$Y_{\rm MSO}$ and $X_{\rm MSO}$--$Z_{\rm MSO}$ planes instead of a single polar plane) and thus optimising the fits.
	\item Because the Martian seasons (monitored by Ls ranges) to a degree and the EUV flux both depend on Mars' heliocentric distance, correlations between these fits are to be expected. For example, similar fits for low Ls values ($<135\deg$) and low EUV flux can be seen in Figure\ \ref{fig:2DFits} (panels c and e, orange curves) and \ref{fig:3DFits} (panels b and c, red curves).
	\item Because the solar cycle is a continuous underlying driver of the shock's position regardless of the binnings adopted here \cite{hall_martian_2019}, we expect also correlations between EUV flux and Mars year results. This effect is most clearly exemplified with the shock standoff decrease when going from the declining phase of solar cycle $24$ (MY32 and MY33, high fluxes) to the next solar minimum (MY34, low fluxes).
\end{itemize}

\section{Conclusions}
In this study, we presented a fast method to estimate automatically the position of the bow shock in real spacecraft orbits, as well as analytical expressions for the normal direction to the shock surface at any point in its close vicinity. After a survey of existing analytical smooth models of the bow shock surface at the planet Mars based on 2D and 3D fits, we used these models as a first prediction of the shock position in the data and refined this prediction further with a predictor-corrector algorithm based on the median absolute deviation of the magnetic field around the predicted shock. This method, biased towards the detection of $q_\perp$ shocks but not entirely limited to them, does not substitute for a detailed analysis of the crossing or for machine-learning techniques currently developed for space missions. It however finds a useful application when it is necessary to quickly determine the position of the spacecraft, or at least an estimate thereof, with respect to the bow shock.

As part of the solar wind and space weather database Helio4Cast, our technique was successfully used to retrieve solar wind undisturbed parameters from the MAVEN mission \cite{mostl_solar_2020}. We also successfully applied the predictor-corrector method to the MAVEN orbit and magnetic field data between November 2014 and February 2021 \cite<see list compiled in>{simon_wedlund_2021}, and performed a series of fits, in 2D and in 3D, to test our method and investigate statistically the shape of the shock depending on Mars Year, solar longitude Ls, and two solar EUV flux levels. The 3D fitting has obvious advantages over the 2D polar axisymmetric geometry usually used to describe the shock structure, namely, a more accurate estimate of asymmetries in the global structure, and taking full advantage of the 3D distribution of bow shock detections in space. This is especially important for bodies such as Mars with large orbital eccentricities and axial tilts to the ecliptic, and for which the heliocentric distance is a strong driver of the EUV flux input and seasonal changes on the planet. 

Expectedly, we found the Martian shock to be highly asymmetric with respect to the North-South hemispheres, in agreement with previous studies \cite<see for example>{hall_annual_2016,halekas_structure_2017,gruesbeck_three-dimensional_2018}. Such an asymmetry is in part linked to the presence of crustal magnetic fields at Mars; however, a specific study taking into account the planet’s rotation and the location of these crustal magnetic sources on the nightside or on the dayside is left for the future. 
Bow shock fits for quasi-perpendicular and parallel shock conditions were, to the precision of our approach, almost identical. In addition, the shock appeared noticeably asymmetric with respect to $Y_{\rm MSO}$ and $Z_{\rm MSO}$ directions in specific conditions, namely, for MY32 and MY35, Ls\,=\,$[135\deg$--$225\deg]$ and larger EUV fluxes. Despite this observed asymmetry, solar-aberrated axisymmetric models may still provide a worthy first approximation of the shock's shape and position. 

To investigate further the conditions of the shock's asymmetry throughout different solar cycles, solar drivers and internal drivers such as crustal magnetic fields, and isolate their respective contribution, a full exploitation of MAVEN's continuously growing datasets is warranted; likewise, a reanalysis of past encounters at Mars using 3D quadric fits would be a welcome addition. These are left for future studies. Applications of these methods, especially in 3D, to other bodies with large orbital eccentricities (such as Mercury) may also prove of interest.

\appendix
\section{Characteristics of a quadric surface}\label{sec:appendix1}

The 3D planetary bow shock in this paper is approximated as a quadratic surface described by the Cartesian equation\,(\ref{eq:quadraticBS}). Mathematically, $17$ different quadrics can exist. However, here only $3$ are physically acceptable for the approximation of a bow shock surface. These are the \emph{`real' ellipsoid}, the \emph{elliptic paraboloid}, and the \emph{hyperboloid of two sheets}. From coefficients $A$ to $I$ defining the quadric's surface equation, it is possible to extract some more `physical' quantities of these surfaces such as the centre of the surface, the direction of the principal axes, the typical length, or the `nose' of the surface. This requires the analysis of one particular matrix $\mathbf{M}$ given by:

\begin{linenomath}
\begin{align}
    \mathbf{M} = \begin{pmatrix}
        A&D/2& F/2\\
        D/2& B& E/2\\
        F/2& E/2& C
        \end{pmatrix}.
\end{align}
\end{linenomath}

Determinant $\text{det}\,\mathbf{M}$ yields useful pieces of information on the considered surface. If $\text{det}\,\mathbf{M}<0$, the surface is an ellipsoid or an hyperboloid of two sheets. If $\text{det}\,\mathbf{M}=0$, it is an elliptic paraboloid. 

The coordinates of the centre of the surface is given by:
\begin{linenomath}
\[
\mathbf{P}_\text{centre}=-\mathbf{M}^{-1}\begin{pmatrix}
        G/2\\
        H/2\\
        I/2
        \end{pmatrix}
\]
\end{linenomath}
if $\mathbf{M}^{-1}$ exists. In the case of an elliptic paraboloid, there is an infinite number of centres placed along the intersections of the two planes of symmetry. 

As $\mathbf{M}$ is symmetric, its eigenvalues are real and eigenvectors are orthogonal. Let us define $\lambda_i~(i=1,2,3)$, the eigenvalues of $\mathbf{M}$, and $\mathcal{V}_i$, their associated eigenvectors. Physically, only $3$ cases should be considered:
\begin{itemize}
    \item ellipsoid: $\lambda_1,\lambda_2,\lambda_3>0$. The characteristic lengths of the ellipsoid are proportional to $1/\sqrt{\lambda_i}$, with the same constant to the length of the conic along the three principal axes of the ellipsoid, 
    \item elliptic paraboloid:  $\lambda_1,\lambda_2>0$ and $\lambda_3=0$,
    \item hyperboloid of two sheets:  $\lambda_1,\lambda_2<0$ and $\lambda_3>0$. The characteristic lengths of the hyperboloid are proportional to $1/\sqrt{|\lambda_i|}$. 
\end{itemize}

Finally, one may be interested in the position of the tip (tail-like direction) or of the nose (subsolar direction) of the surface. These extremum points are at a distance $L_1$, $L_2$, and $L_3$ from the centre in the direction $\pm\mathcal{V}_i$. Therefore they are given by:
\begin{linenomath}
\[L_i=\left(\sqrt{\lambda_i}\right)^{-1} \sqrt{\mathbf{P}_\text{centre}^{T}\,\mathbf{M}\,\mathbf{P}_\text{centre}+1}\]
\end{linenomath}
and 
\begin{linenomath}
\begin{align}
\mathbf{P}_{\pm,\text{ext}}=\mathbf{P}_\text{centre}
        \begin{pmatrix}
        1&1&1
        \end{pmatrix}
        \pm
        \begin{pmatrix}
           L_1\mathcal{V}_1 &L_2\mathcal{V}_2 &L_3\mathcal{V}_3
        \end{pmatrix}
        \label{eq:noseEllipsoid}
\end{align}
\end{linenomath}
where the columns of $\mathbf{P}_{\pm,\text{ext}}$ are the locations of the extrema. For a hyperboloid of two sheets, only the real solution associated with $\lambda_3$ should be considered: this gives the position of the noses or tips of both sheets. The sunward-most position of the ellipsoid (its nose) is referred to as $\mathbf{P}_\text{nose}$.

The tip or nose of the ellipsoid is in our context along the direction of the eigenvectors with the largest $X$ (in absolute value) component. 

The volume of the ellipsoid is:
\begin{linenomath}
\begin{align}
    \mathrm{V} = \frac{4}{3}\pi\ L_1\,L_2\,L_3.
\end{align}
\end{linenomath}

Table\ \ref{tab:Conic3Dphysics} presents the length of each of the three principal axes of the quadric $L_i$, the ellipsoid's volume, eigenvectors $\mathcal{V}_i$ and the coordinates of the centre and sunward nose of the surface for each ellipsoid in Table\ \ref{tab:Conic3Dparameters}.

\begin{sidewaystable}
    \centering
    \caption{Characteristics of the 3D Martian bow shock as derived from the MAVEN dataset, see Table\ \ref{tab:Conic3Dparameters} for the parameters of the 3D surfaces considered. All quadrics are ellipsoids. $L_i$ ($i=1,2,3$) are the lengths in units of $R_p$ of the principal axes of the ellipsoids and $\mathrm{V}$ their volume. $\mathcal{V}_i$ are the eigenvectors of matrix $\mathbf{M}$, i.e. the normalised directions of the principal axes in MSO Cartesian coordinates (because the values are normalised to $R_p$ and rounded down, a value of 1.00 or 0.00 is not stricto sensu 1 or 0). $\mathbf{P}_\text{centre}$ and $\mathbf{P}_\text{nose}$ are the positions of the centre of the ellipsoid and its sunward nose, in MSO Cartesian coordinates. The domain of validity for each fit is shown in Figure\ \ref{fig:3DFits}: fits are valid for $X_{\rm MSO}\geq-0.5\,R_p$ on average. Mars' radius is $R_\text{p} = 3,389.5$\,km.
    }
    \scriptsize
    \begin{tabular}{l@{\hskip 0.1in} c@{\hskip 0.1in} c @{\hskip 0.1in}c | c | r r r | r r r | r r r | r r r | r r r}
        {\bf Case} & $L_1$ & $L_2$ & $L_3$  & Volume $\mathrm{V}$ & \multicolumn{3}{c}{$\mathcal{V}_1$ [$R_p$]} & \multicolumn{3}{c}{$\mathcal{V}_2$ [$R_p$]} & \multicolumn{3}{c}{$\mathcal{V}_3$ [$R_p$]} & \multicolumn{3}{c}{$\mathbf{P}_{\rm centre}$ $[R_p]$}& \multicolumn{3}{c}{$\mathbf{P}_{\rm nose}$ $[R_p]$}\\
        &  &  $[R_p]$ & & $[R_p^3]$ & $X_{\rm MSO}$ & $Y_{\rm MSO}$ & $Z_{\rm MSO}$ & $X_{\rm MSO}$ & $Y_{\rm MSO}$ & $Z_{\rm MSO}$ & $X_{\rm MSO}$ & $Y_{\rm MSO}$ & $Z_{\rm MSO}$ & $X_{\rm MSO}$ & $Y_{\rm MSO}$ & $Z_{\rm MSO}$ & $X_{\rm MSO}$ & $Y_{\rm MSO}$ & $Z_{\rm MSO}$\\
         \hline
     Gruesbeck+ (2018), all points & 8.20 & 4.34 & 4.11 & 613.30 & -0.97 & 0.10 & 0.22 & 0.08 & -0.72 & 0.69 & 0.22 & 0.69 & 0.69 & -6.45 & 0.59 & 0.99 & 1.52 & -0.21 & -0.80 \\
     All points, this work & 2.84 & 2.74 & 2.55 & 83.37 & -0.44 & -0.06 & 0.90 & -0.16 & 0.99 & -0.01 & -0.88 & -0.15 & -0.45 & -1.07 & 0.12 & 0.05 & 1.18 & 0.50 & 1.19 \\
     MY32, this work & 3.42 & 3.09 & 2.68 & 118.38 & 0.76 & -0.28 & -0.59 & 0.13 & -0.82 & 0.55 & 0.64 & 0.50 & 0.59 & -1.44 & 0.17 & 0.17 & 1.16 & -0.79 & -1.83 \\
     MY33, this work & 2.97 & 2.82 & 2.65 & 92.83 & -0.50 & 0.61 & 0.62 & -0.06 & 0.69 & -0.72 & -0.87 & -0.40 & -0.31 & -1.17 & 0.19 & 0.09 & 1.13 & 1.23 & 0.90 \\
     MY34, this work & 2.85 & 2.70 & 2.60 & 83.79 & -0.37 & 0.05 & 0.93 & -0.65 & -0.73 & -0.21 & 0.67 & -0.68 & 0.30 & -1.15 & 0.04 & 0.02 & 0.60 & -1.73 & 0.80 \\
     MY35, this work & 2.28 & 2.13 & 1.34 & 27.19 & -0.13 & 0.06 & 0.99 & 0.08 & 1.00 & -0.05 & -0.99 & 0.08 & -0.13 & 0.13 & 0.01 & 0.00 & 1.45 & -0.09 & 0.17 \\
     Ls = [$315^\circ-45^\circ$], this work & 2.99 & 2.86 & 2.73 & 97.79 & 0.78 & 0.62 & 0.12 & -0.25 & 0.13 & 0.96 & 0.58 & -0.78 & 0.25 & -1.38 & -0.12 & -0.01 & 0.94 & 1.73 & 0.34 \\
     Ls = [$45^\circ-135^\circ$], this work & 3.08 & 2.68 & 2.52 & 87.10 & -0.61 & -0.23 & 0.76 & 0.21 & 0.87 & 0.44 & -0.76 & 0.43 & -0.48 & -1.23 & -0.02 & 0.20 & 0.69 & -1.10 & 1.42 \\
     Ls = [$135^\circ-225^\circ$], this work & 2.80 & 2.36 & 1.88 & 51.95 & 0.13 & -0.99 & 0.11 & 0.74 & 0.03 & -0.67 & 0.65 & 0.17 & 0.74 & -0.59 & 0.35 & 0.38 & 1.17 & 0.41 & -1.19 \\
     Ls = [$225^\circ-315^\circ$], this work & 3.13 & 2.86 & 2.65 & 99.65 & -0.48 & -0.33 & 0.81 & -0.47 & 0.88 & 0.08 & -0.74 & -0.34 & -0.58 & -1.18 & 0.02 & 0.24 & 0.79 & 0.94 & 1.76 \\
     EUV flux $\geq$ 0.0028\,W/m$^2$ & 3.81 & 3.13 & 3.00 & 150.08 & -0.92 & 0.25 & 0.32 & -0.02 & -0.81 & 0.58 & -0.40 & -0.53 & -0.75 & -2.05 & 0.29 & 0.22 & 1.44 & -0.67 & -0.98 \\
     EUV flux $< 0.0028$\,W/m$^2$ & 2.73 & 2.54 & 2.27 & 65.77 & -0.26 & -0.02 & 0.97 & -0.27 & -0.96 & -0.09 & 0.93 & -0.28 & 0.24 & -0.81 & 0.01 & 0.02 & 1.30 & -0.63 & 0.57 \\
     Quasi-$\perp$ & 2.87 & 2.73 & 2.53 & 83.11 & -0.43 & -0.10 & 0.90 & -0.09 & 0.99 & 0.06 & -0.90 & -0.06 & -0.44 & -1.06 & 0.11 & 0.07 & 1.21 & 0.26 & 1.18 \\
     Quasi-$\parallel$ & 3.00 & 2.79 & 2.75 & 96.12 & -0.99 & -0.02 & 0.10 & 0.09 & -0.66 & 0.75 & 0.06 & 0.75 & 0.66 & -1.40 & 0.03 & -0.05 & 1.58 & 0.08 & -0.36 \\
        \hline   
    \end{tabular}
    \label{tab:Conic3Dphysics}
\end{sidewaystable}

\section{Subsolar tip of the trace of an ellipsoid surface in Cartesian coordinates}\label{sec:appendix2}

The subsolar point of the projection of a 3D ellipsoid in 2D planes, as shown in Figure\ \ref{fig:3DFits} (crosses), can be obtained by finding the roots of the corresponding 2D conic in the plane considered.
For $z=0$, Equation\,(\ref{eq:quadraticBS}) becomes a second order equation:
\begin{linenomath}
\begin{align}
     A x^2 + B y^2 + D xy + G x + Hy - 1 = 0. 
\end{align}{}
\end{linenomath}
Fixing variable $y$, the equation can be put in quadratic form with the following positive root:
\begin{linenomath}
\begin{align}
     x_{\rm M} &= \frac{-(Dy+G)+\sqrt{\Delta}}{2A},\label{eq:quadraticBSxy}\\
     \Delta &= (Dy+G)^2 - 4A\left(By^2+Hy-1\right) > 0
\end{align}{}
\end{linenomath}
Finding the maximum of this function is equivalent to finding a $y$ value that maximises this function. Posing $\xi = Dy+G$, its derivative has the form:
\begin{linenomath}
\begin{align}
	\pd{x_{\rm M}}{y}= \frac{2D \xi - 4A (2By+H)}{4A\sqrt{\xi^2 - 4A\left(By^2+Hy-1\right)}} - \frac{D}{2A}
\end{align}
\end{linenomath}
Solving $\pd{x_{\rm M}}{y}=0$ for $y$ and using that result in Equation\,(\ref{eq:quadraticBSxy}) makes it possible to calculate the final $(x,y)$ coordinates of the projected ellipsoid's tip in the corresponding $x-y$ plane. An identical reasoning can be made for the $x-z$ plane.

This tip in a plane is however not necessarily the farthest subsolar point of the ellipsoid's surface. Its position in 3D is by contrast given by Equation\,(\ref{eq:noseEllipsoid}) in \ref{sec:appendix1}.

%
%
%
%
%
%
%
%

\acknowledgments
C. Simon Wedlund, M. Volwerk and C. Möstl thank the Austrian Science Fund (FWF): P32035-N36, P31659-N27, P31521-N27. A. Beth thanks the Swedish National Space Agency (SNSA) and its support with the grant 108/18. Parts of this work for the observations obtained with the SWEA
instrument are supported by the French space agency CNES. CSW thanks M. Simon Wedlund for insightful comments and discussions. The authors acknowledge Emmanuel Penou for help and access to the CLWeb software (v16.09) from IRAP/Observatoire Midi-Pyrénées. The authors thank Dr. Jared Espley (NASA) for his support in using MAVEN/MAG data. 
CSW acknowledges L. Hunyadi for the matrix implementation of the Taubin 3D fitting algorithm, and Yair Altman for developing and maintaining the Matlab package ``export\_fig'' for figure pdf exports. Finally, the authors would like to acknowledge ISSI for the opportunity it offered for very valuable discussions on this topic as part of the International Team $\#499$ ``Similarities and Differences in the Plasma at Comets and Mars'' led by C. Götz during these hard Covid times.

{\bf Data Availability Statement.}
The calibrated MAVEN/MAG datasets are freely available from the NASA Planetary Data System (PDS) at \url{https://doi.org/10.17189/1414178}. 
The corresponding predicted bow shock times and spatial coordinates for the 2014--2021 dataset using our predictor-corrector algorithm are provided for reference on Zenodo at \url{https://doi.org/10.5281/zenodo.5725288} \cite<Version 2, >{simon_wedlund_2021}.
The FISM-P Mars Solar Spectral Irradiance model is available at \url{https://lasp.colorado.edu/lisird/data/fism_p_ssi_mars/}.
The Helio4Cast database is available at \url{www.helioforecast.space/icmecat} and \url{www.helioforecast.space/sircat}. The solar wind monitor dataset at Mars was specifically derived for Helio4Cast using our predictor algorithm and can be downloaded at \url{https://doi.org/10.6084/m9.figshare.6356420}.



%
%

\bibliography{bibliography} 

%
%
%
%
%

\end{document}